\documentclass[aps,pra,showpacs,twocolumn]{revtex4}
\usepackage{hyperref}
\usepackage{amsmath}
\usepackage{amssymb}
\usepackage{amsthm}
\usepackage{mathrsfs} 

\usepackage{graphicx}
\usepackage{xcolor}
\newcommand\mb{\boldsymbol}

\begin{document}

\title{Resolution of two fundamental issues in the dynamics of relativity and \\ exposure of a real version of the emperor's new clothes}
\author{Changbiao Wang }
\email{changbiao\_wang@yahoo.com}
\affiliation{ShangGang Group, 10 Dover Road, North Haven, CT 06473, USA}

\begin{abstract}
In this paper, we aim to resolve two fundamental issues in the dynamics of relativity: (i) Under what condition, the time-column space integrals of a Lorentz four-tensor constitute a Lorentz four-vector, and (ii) under what condition, the time-element space integral of a Lorentz four-vector is a Lorentz scalar; namely two ``conservation laws'', which are mispresented in traditional textbooks, and widely used in fundamental research, such as relativistic analysis of the momentum of light in a medium, and the proofs of the positive mass theorem in general relativity.  To resolve issue (i), we have developed a generalized Lorentz \emph{four-vector} theorem based on the principles of classical mathematical analysis, with a simplified analytic example given to illustrate how to transform a space integral from one inertial frame to another, and a strict mathematical derivation provided to confirm the effect of Lorentz contraction.  We use this four-vector theorem to verify M\o ller's theorem, and surprisingly find that M\o ller's theorem is fundamentally wrong.  We provide a corrected version of M\o ller's theorem.  We also use this four-vector theorem to analyze a plane light wave in a moving uniform medium, and find that the momentum and energy of Minkowski quasi-photon constitute a Lorentz four-vector and Planck constant is a Lorentz invariant.  To resolve issue (ii), we have developed a generalized Lorentz \emph{scalar} theorem.  We use this theorem to verify the ``invariant conservation law'' in relativistic electrodynamics, and unexpectedly find that it is also fundamentally wrong. Thus the two ``conservation laws'' in traditional textbooks, which have magically attracted several generations of most outstanding scientists, turned out to be imaginary, just like the emperor's new clothes; creating a scientific myth in the modern theoretical and mathematical physics: Believing is seeing.
\end{abstract} 

\maketitle 

\section{Introduction}
\label{s1}
In the dynamics of relativity, the energy and momentum of a physical system is described by a Lorentz four-tensor; such a tensor is usually called \emph{energy--momentum tensor} \cite{r1,r2,r3}, \emph{stress tensor} \cite{r4}, \emph{stress--energy tensor} \cite{r5,r6}, or \emph{momentum--energy stress tensor} \cite{r7}.  If the tensor is divergence-less, then the system is thought to be conserved \cite{r1,r2,r3}, and it is a closed system \cite{r3}; thus the total energy and momentum can be obtained by carrying out space integration of the time-column elements of the tensor to constitute a Lorentz four-vector \cite{r1,r2,r3,r4,r5,r8,r9,r10,r11}. 

Mathematically speaking, if a tensor satisfies certain conditions, the space integrals of the tensor's time-column elements can form a Lorentz four-vector.  For the sake of convenience, we call such a mathematical statement ``four-vector theorem''.

Laue set up a four-vector theorem for a tensor that is required to be time-independent \cite{r12}.  Laue's theorem only provides a sufficient condition (instead of a sufficient and necessary condition), and it cannot be used to judge the Lorentz property of the energy and momentum of electrostatic fields.  In a recent study, Laue's theorem is improved to be a theorem that has a sufficient and necessary condition, and it is successfully used to generally resolve the electrostatic field problem \cite{r6}.

In contrast to Laue's theorem, M\o ller provided a four-vector theorem for a tensor that is required to be divergence-less, with a boundary condition imposed, but allowed to be time-dependent \cite{r3}.  M\o ller's theorem only has a sufficient condition (instead of a sufficient and necessary condition) \cite{r13}, but it is more attractive because the energy--momentum tensor for electromagnetic (EM) radiation fields varies with time \cite{r8,r9,r10,r11}.  It is widely recognized in the community that M\o ller's theorem is \emph{absolutely rigorous} so that this theorem has been widely used in quantum electrodynamics \cite{r8} and relativistic analysis of light momentum in a dielectric medium \cite{r9,r10,r11}.

In this paper, we provide a generalized Lorentz four-vector theorem for a tensor which is not required to be time-independent and divergence-less, and on which no boundary conditions are imposed.  This theorem has a sufficient and necessary condition.  We use this theorem to verify M\o ller's theorem, surprisingly finding that M\o ller's theorem is fundamentally wrong.

Like the four-vector theorem, a Lorentz \emph{scalar} theorem is a mathematical statement that under what conditions, the time-element space integral of a four-vector is a Lorentz scalar.  In Ref.\,\cite{r6}, a scalar theorem for a four-vector that is required to be time-independent is set up, called ``derivative von Laue's theorem'', and it is successfully used to strictly resolve the invariance problem of total electric charge in relativistic electrodynamics.

In this paper, we also provide a generalized Lorentz scalar theorem for a four-vector which is not required to be time-independent and divergence-less, and on which no boundary conditions are imposed.  This theorem has a sufficient and necessary condition.  We use this scalar theorem to identify the validity of a well-known result in the dynamics of relativity that if a Lorentz four-vector is divergence-less, then the time-element space integral of the four-vector is a Lorentz scalar \cite[p.\,168]{r3}, namely the ``invariant conservation law'' claimed by Weinberg \cite[p.\,41]{r2}. We unexpectedly find that this widely-accepted result is also fundamentally wrong.

The paper is organized as follows.  In Sec.\,\ref{s2}, proofs are given of Lorentz four-vector and scalar theorems based on the principles of classical mathematical analysis, with a simplified analytic example given to illustrate how to transform a space integral from one inertial frame to another, and a strict mathematical derivation provided to confirm the effect of Lorentz contraction.  In Sec.\,\ref{s3}, M\o ller's theorem is proved to be incorrect, and a corrected version of M\o ller's theorem is provided.  In Sec.\,\ref{s4}, the ``invariant conservation law'' in relativistic electrodynamics is proved to be invalid; namely the current continuity equation $\nabla\cdot\mathbf{J}+\partial\rho/\partial t=0$ cannot be taken as the charge conservation law in relativity.  In Sec.\,\ref{s5}, some remarks and conclusions are given. In Appendix \ref{appa} it is demonstrated why the hyperplane differential-element four-vector, introduced to transform space integrals between Lorentz inertial frames in textbooks, contradicts the principles of mathematical analysis and the principle of relativity.  In  Appendix \ref{appb}, as an application of Theorem 1 to Minkowski tensor for a plane light wave in a moving uniform medium, the momentum--energy four-vector of the quasi-photon and the Lorentz invariance of Planck constant are naturally derived.  In Appendix \ref{appc}, physical counterexamples of Thirring's claims are provided. In Appendix \ref{appd}, an illustration is given of why the proofs of the positive mass theorem in general relativity are based on a flawed theoretical framework. In Appendix \ref{appe}, the covariance of Gordon optical metric is questioned.

\section{Lorentz four-vector theorems and scalar theorem}
\label{s2}
In this section, proofs are given of Lorentz four-vector and scalar theorems, and a simplified analytic example is provided to illustrate how to transform a space integral from one inertial frame to another and to explain why under time-space Lorentz transformation the effect of Lorentz contraction, namely the relativistic effect of lengths of a rigid rod argued by Einstein according to the principle of relativity \cite{r20}, is strictly supported by the principles of classical mathematical analysis.    

Four-vector theorems provide a criterion to judge under what condition the space integrals of the time-\emph{column} elements of a tensor constitute a Lorentz four-vector (Theorem 1) and under what condition the space integrals of the time-\emph{row} elements of a tensor constitute a Lorentz four-vector (Theorem 2), while the scalar theorem provides a criterion to judge under what condition the space integral of the time-element of a four-vector is a Lorentz scalar (Theorem 3).  The proofs of Theorem 1 and Theorem 2 are very similar, and without loss of generality, only the proof of Theorem 1 is given.

Suppose that an inertial frame of $X'Y'Z'$ moves uniformly at $\mb{\beta}c$ relatively to the laboratory frame $XYZ$, where $c$ is the vacuum light speed.  The Lorentz transformation of time-space four-vector  $X^{\mu}=(\textbf{x},ct)$ is given by \cite{r6,r7} 
\begin{align}
&\mathbf{x}'=\mathbf{x}+\xi(\mb{\beta}\cdot\mathbf{x})\mb{\beta}-\gamma\mb{\beta}ct,  
\label{eq1} \\ 
&ct'=\gamma(ct-\mb{\beta}\cdot\mathbf{x}),   
\label{eq2} 
\end{align} 
or conversely, given by 
\begin{align}
&\mathbf{x}=\mathbf{x}'+\xi(\mb{\beta}'\cdot\mathbf{x}')\mb{\beta}'-\gamma\mb{\beta}'ct',  
\label{eq3} \\  
&ct=\gamma(ct'-\mb{\beta}'\cdot\mathbf{x}'),   
\label{eq4}
\end{align}  
where $\xi\equiv (\gamma-1)/\mb{\beta}^{2}=\gamma^2/(\gamma+1)$, $\gamma\equiv (1-\mb{\beta}^2)^{-1/2}$, and $\mb{\beta}'=-\mb{\beta}$. Note: $X_{\mu}=g_{\mu\nu}X^{\nu}=(-\mathbf{x},ct)$, with $g_{\mu\nu}=g^{\mu\nu}=\mathrm{diag}(-1,-1,-1,+1)$ the Minkowski metric.

According to the definition of tensors \cite[p.108]{r3}, if $\Omega^{\mu\nu}(\mathbf{x},t)$ is a Lorentz four-tensor given in $XYZ$, where $\mu,\nu$ = 1, 2, 3, and 4, with the index 4 corresponding to time component, then in $X'Y'Z'$  the tensor $\Omega'^{\mu\nu}\big(\mathbf{x}=\mathbf{x}(\mathbf{x}',t'),t=t(\mathbf{x}',t')\big)$  is obtained through ``double'' Lorentz transformation of $\Omega^{\mu\nu}(\mathbf{x},t)$, given by \\
\begin{align}
& \Omega'^{\mu\nu}(\mathbf{x},t)=\frac{\partial X'^{\mu}}{\partial X^{\lambda}}\frac{\partial X'^{\nu}}{\partial X^{\sigma}}\Omega^{\lambda\sigma}(\mathbf{x},t),  
\label{eq5}
\\
& \Omega'^{\mu\nu}\Big(\mathbf{x}=\mathbf{x}(\mathbf{x}',t'),t=t(\mathbf{x}',t')\Big)
\notag \\
& \hspace{5mm} =\frac{\partial X'^{\mu}}{\partial X^{\lambda}}\frac{\partial X'^{\nu}}{\partial X^{\sigma}}   
\Omega^{\lambda\sigma}\Big(\mathbf{x}=\mathbf{x}(\mathbf{x}',t'),t=t(\mathbf{x}',t')\Big),   
\label{eq6} 
\end{align} \\
where $\partial X'^{\mu}/\partial X^{\lambda}$ and $\partial X'^{\nu}/\partial X^{\sigma}$ are obtained from Lorentz transformation Eqs.\,(\ref{eq1}) and (\ref{eq2}), while $\mathbf{x}=\mathbf{x}(\mathbf{x}',t')$ and $t=t(\mathbf{x}',t')$ denote Lorentz transformation Eqs.\,(\ref{eq3}) and (\ref{eq4}), respectively.  Eq.\,(\ref{eq5}) is the expression of $\Omega'^{\mu\nu}$ observed in $XYZ$, and Eq.\,(\ref{eq6}) is the expression of $\Omega'^{\mu\nu}$ observed in $X'Y'Z'$. \vspace{5mm}

\noindent \textbf{Theorem 1.} Suppose that $\Theta^{\mu\nu}(\mathbf{x},t)$  is an integrable Lorentz four-tensor, defined in the domain $V$ in the laboratory frame $XYZ$, where $\mu, \nu$ = 1, 2, 3, and 4, with the index 4 corresponding to time component, and $V$ including its boundary is at rest in $XYZ$, namely any $\mathbf{x}\in V$ is independent of $t$.  The space integrals of the time-\emph{column} elements of the tensor in $XYZ$ are defined as 
\begin{equation}
P^{\mu}=\int_{V:~t=const}\Theta^{\mu 4}(\mathbf{x},t)\mathrm{d}^3x.
\label{eq7}
\end{equation} \\
The space integrals of time-\emph{column} elements of the tensor in $X'Y'Z'$ are defined as  
\begin{equation}
P'^{\mu}=\int_{V':~t,\,t'=const}\Theta'^{\mu 4}\Big(\mathbf{x}=\mathbf{x}(\mathbf{x}',t'),t\Big)\mathrm{d}^3x',
\label{eq8}
\end{equation} 
where
\begin{align}
\Theta'^{\mu 4}&\Big(\mathbf{x}=\mathbf{x}(\mathbf{x}',t'),t\Big)
\notag \\
&:=\frac{\partial X'^{\mu}}{\partial X^{\lambda}}\frac{\partial X'^{4}}{\partial X^{\sigma}}\Theta^{\lambda\sigma}\Big(\mathbf{x}=\mathbf{x}(\mathbf{x}',t'),t\Big).
\label{eq9}
\end{align} \\
The four-vector theorem states: $P^{\mu}$ is a Lorentz four-vector \emph{if and only if} 
\begin{align}
\int_{V:~t=const}&\Theta^{\mu j}(\mathbf{x},t)\mathrm{d}^3x=0 
\notag \\
\quad&\mathrm{for} \quad\mu=1,2,3,4 \quad\mathrm{and}\quad j=1,2,3
\label{eq10}
\end{align}
holds. \vspace{4mm}

There are a few main points to understand Theorem 1 that should be noted, as follows. \vspace{2mm}

\indent (i) The importance of the definition Eq.\,(\ref{eq9}) should be emphasized, otherwise the implication of $P'^{\mu}=\int\Theta'^{\mu 4}\mathrm{d}^3x'$ is ambiguous, and we cannot set up the transformation between $P^{\mu}$ and $P'^{\mu}$.  In Eq.\,(\ref{eq9}), the space variables $\mathbf{x}$ in $\Theta^{\lambda\sigma}(\mathbf{x},t)$ are replaced by  $\mathbf{x}=\mathbf{x}(\mathbf{x}',t')$, namely the space Lorentz transformation Eq.\,(\ref{eq3}), while $t$ in $\Theta^{\lambda\sigma}(\mathbf{x},t)$ is kept as it is.  Note that in  Eq.\,(\ref{eq7}) for the definition of  $P^{\mu}$, the integration variables $\mathbf{x}=(x,y,z)$ are \textit{independent of} $t$ because the domain $V$ is fixed in  $XYZ$, which is the mathematical reason why $\mathbf{x}$ in $\Theta^{\lambda\sigma}(\mathbf{x},t)$ in Eq.\,(\ref{eq9}) should be replaced by  $\mathbf{x}=\mathbf{x}(\mathbf{x}',t')$ [Eq.\,(\ref{eq3})], instead of $\mathbf{x}=f(\mathbf{x}',t)$ derived from $\mathbf{x}'=\mathbf{x}'(\mathbf{x},t)$  given by Eq.\,(\ref{eq1}) where $\mathbf{x}$  are also \textit{functions of} $t$  in addition to  $\mathbf{x}'$.\vspace{2mm}

\indent (ii) Observed in  $XYZ$, like $P^{\mu}$,  $P'^{\mu}$  is \emph{only dependent on} $t$ in general; confer Eq.\,(\ref{eq16}).  The quantity $t'$  in the integrand $\Theta'^{\mu 4}\big(\mathbf{x}=\mathbf{x}(\mathbf{x}',t'),t\big)$  of $P'^{\mu}=P'^{\mu}(t)=\int_{V':~t,\,t'=const}\Theta'^{\mu 4}\big(\mathbf{x}=\mathbf{x}(\mathbf{x}',t'),t\big)\mathrm{d}^3x'$ is introduced as a constant parameter in the space integral transform from $XYZ$ to $X'Y'Z'$, and thus observed in $X'Y'Z'$, the boundary of $V'$ is moving so that $P'^{\mu}=P'^{\mu}(t)$ does not contain $t'$.  \vspace{2mm}

\indent (iii) If $\Theta^{\lambda\sigma}(\mathbf{x},t)$ is independent of $t$, then both $P^{\mu}$ and $P'^{\mu}$ are independent of $t$, namely they are constants.  \vspace{2mm}

\indent (iv) The symmetry ($\Theta^{\mu\nu}=\Theta^{\nu\mu}$) and divergence-less ($\partial_{\nu}\Theta^{\mu\nu}$=0) are not required, and there are no boundary conditions imposed on $\Theta^{\mu\nu}(\mathbf{x},t)$. \vspace{4mm}

\noindent \emph{Analytical example.}  In order to better understand (i) and (ii), let us take a simple one-dimensional example to illustrate how to transform the space integral from one Lorentz inertial frame to another according to the principles of classical mathematical analysis \cite{r14}.

Suppose that $X'Y'Z'$  moves at $|\mb{\beta}|c$  with respect to $XYZ$  along the positive $x$-direction. In such a case, the space Lorentz transformations Eqs.\,(\ref{eq1}) and (\ref{eq3}) are, respectively, simplified into:  
\begin{align}
&\hspace{2mm} x'=\gamma(x-|\mb{\beta}|ct), &y'=y, &\quad z'=z \quad\mathrm{for ~Eq.\,(\ref{eq1}),}
\notag \\
&\hspace{2mm} x=\gamma(x'+|\mb{\beta}|ct'), &y=y', &\quad z=z' \quad\mathrm{for ~Eq.\,(\ref{eq3}).} 
\notag
\end{align} 

Consider the space integral transform from $XYZ$  to  $X'Y'Z'$, given by 
\begin{align}
I(t)&=\int_{a}^{b}\cos(x-ct)\mathrm{d}x
\notag \\
&=\int_{~x'_a~[=a/\gamma-|\mb{\beta}|ct']}^{~x'_b~[=b/\gamma-|\mb{\beta}|ct']}\cos\Big(\gamma(x'+|\mb{\beta}|ct')-ct\,\Big)\gamma\mathrm{d}x' 
\notag \\
\Big[&=\sin(b-ct)-\sin(a-ct)\,\,\Big], \nonumber 
\end{align} 
where the integration region $[a,b]$ is fixed in $XYZ$, and any $x\in [a,b]$ is independent of $t$, while observed in  $X'Y'Z'$, according to the principle of relativity, the corresponding integration region $[x'_a,x'_b]$ must move at $|\mb{\beta}|c$  along the minus $x$-direction.   

In the above integral transform, a key problem to be solved is to determine which Lorentz transformation should be taken, Eq.\,(\ref{eq1}) or Eq.\,(\ref{eq3}), as shown below. 
\begin{enumerate}
\item[(a)] \emph{Why Eq.\,(\ref{eq3}) is taken?} From $\int_{a}^{b}\cos(x-ct)\mathrm{d}x$ in $XYZ$ to  $\int_{x'_a}^{x'_b}\cos\big(\gamma(x'+|\mb{\beta}|ct')-ct\,\big)\gamma\mathrm{d}x'$ in $X'Y'Z'$, $x$ in $\int_{a}^{b}\cos(x-ct)\mathrm{d}x$ must be replaced by \vspace{1mm}
 
\hspace{5mm} $x=\gamma(x'+|\mb{\beta}|ct')$  \quad [Eq.\,(\ref{eq3})], \vspace{1mm}

instead of \vspace{1mm}

\hspace{5mm} $x=x'/\gamma+|\mb{\beta}|ct$\quad\quad derived from 

\hspace{5mm} $x'=\gamma(x-|\mb{\beta}|ct)$ ~\quad [Eq.\,(\ref{eq1})], \vspace{1mm}

because the (proper) integration region $a\le x\le b$  is fixed in $XYZ$, and the region boundaries $x=a$ and $x=b$  are \emph{independent of} $t$, while $x=x'/\gamma+|\mb{\beta}|ct$ derived from $x'=\gamma(x-|\mb{\beta}|ct)$ [Eq.\,(\ref{eq1})] is a \emph{function of} $t$ in addition to $x'$.

\item[(b)] \emph{Differential element transformation.}  According to (a) which is a strict \emph{mathematical rule}, the differential element transformation must be calculated from $x=\gamma(x'+|\mb{\beta}|ct')$  [Eq.\,(\ref{eq3})], with $t'$  taken as a constant parameter, leading to $\mathrm{d}x=\gamma\mathrm{d}x'$ where the Jacobian determinant $\partial (x)/\partial (x')=\partial x/\partial x'=\gamma$  is taken into account. 

\item[(c)] \emph{Motion of region.}  Observed in  $X'Y'Z'$, the integration region is given by $x'_a\le x'\le x'_b$, and the region boundaries $x'_a ~[=a/\gamma-|\mb{\beta}|ct']$ and  $x'_b ~[=b/\gamma-|\mb{\beta}|ct']$ are moving at a velocity of $|\mb{\beta}|c$ along the minus $x$- or $x'$-direction  so that the integral does not contain $t'$  although the integrand $\cos\big(\gamma(x'+|\mb{\beta}|ct')-ct\,\big)\gamma$ contains  $t'$.  

\item[(d)] \emph{Effect of Lorentz contraction.} The expressions of integration region boundaries \vspace{1mm}
 
\hspace{15mm} $x'_a=a/\gamma-|\mb{\beta}|ct'$ 
  
\hspace{15mm} $x'_b=b/\gamma-|\mb{\beta}|ct'$ \vspace{1mm}

are governed by the mathematical rule (a), suggesting that  $x'_a$  and  $x'_b$  must be ``measured'' at the same time  $t'$ physically, except for the case with  $\mb{\beta}c=0$ $\Rightarrow$ $X'Y'Z'$ coincides with $XYZ$, and $x'_a~(=a)$  and  $x'_b~(=b)$ are independent of the times when they are measured, respectively.  Thus the effect of Lorentz contraction or the relativistic effect of lengths of a rigid rod, defined by \vspace{1mm} 

\hspace{10mm} $\int_{a}^{b}\mathrm{d}x=\int_{x'_a}^{x'_b}(\gamma\mathrm{d}x')=\gamma\int_{x'_a}^{x'_b}\mathrm{d}x'$ \vspace{1mm}

\hspace{11mm} $\Longleftrightarrow$ \hspace{5mm} $x'_b-x'_a=(b-a)/\gamma$, \vspace{1mm}

is a strict and natural result from the principles of mathematical analysis, where $\int_{a}^{b}\mathrm{d}x$ is the proper length of the rod, and $\int_{x'_a}^{x'_b}\mathrm{d}x'$  is exactly the length of the moving rod defined by Einstein, because  $x'_a$  and  $x'_b$  in  $\int_{x'_a}^{x'_b}\mathrm{d}x'$ are the points at which ``the two ends of the [moving] rod to be measured are located \emph{at a definite time} [\,at the same time $t'$\,]'' \cite{r20}. \vspace{2mm}
\end{enumerate}

\noindent \textbf{Theorem 2.} Suppose that $\Theta^{\mu\nu}(\mathbf{x},t)$  is an integrable Lorentz four-tensor, defined in the domain $V$ in the laboratory frame $XYZ$, where $\mu, \nu$ = 1, 2, 3, and 4, with the index 4 corresponding to time component, and $V$ including its boundary is at rest in $XYZ$, namely any $\mathbf{x}\in V$ is independent of $t$.  The space integrals of the time-\emph{row} elements of the tensor in $XYZ$ are defined as 
\begin{equation}
\mathnormal{\Pi}^{\nu}=\int_{V:~t=const}\Theta^{4\nu}(\mathbf{x},t)\mathrm{d}^3x.
\label{eq11}
\end{equation} 
The space integrals of time-\emph{row} elements of the tensor in $X'Y'Z'$ are defined as 
\begin{equation}
\mathnormal{\Pi}'^{\nu}=\int_{V':~t,\,t'=const}\Theta'^{4\nu}\Big(\mathbf{x}=\mathbf{x}(\mathbf{x}',t'),t\Big)\mathrm{d}^3x',
\label{eq12}
\end{equation} 
where
\begin{align}
\Theta'^{4\nu}&\Big(\mathbf{x}=\mathbf{x}(\mathbf{x}',t'),t\Big)
\notag \\
&:= \frac{\partial X'^{4}}{\partial X^{\lambda}}\frac{\partial X'^{\nu}}{\partial X^{\sigma}}\Theta^{\lambda\sigma}\Big(\mathbf{x}=\mathbf{x}(\mathbf{x}',t'),t\Big).
\label{eq13}
\end{align} 
The four-vector theorem states: $\mathnormal{\Pi}^{\nu}$ is a Lorentz four-vector \emph{if and only if} 
\begin{align}
\int_{V:~t=const}&\Theta^{i\nu}(\mathbf{x},t)\mathrm{d}^3x=0
\notag \\ 
\quad&\mathrm{for} \quad\nu=1,2,3,4 \quad\mathrm{and}\quad i=1,2,3
\label{eq14}
\end{align} 
holds.\\ 

\noindent \textbf{Proof of Theorem 1.} From Eqs.\,(\ref{eq8}) and (\ref{eq9}) we have \\
\begin{align}
&P'^{\mu}=\int_{V':~t,\,t'=const}\Theta'^{\mu 4}\Big(\mathbf{x}=\mathbf{x}(\mathbf{x}',t'),t\Big)\mathrm{d}^3x'
\notag \\
~~\notag \\
&=\frac{\partial X'^{\mu}}{\partial X^{\lambda}}\frac{\partial X'^{4}}{\partial X^{\sigma}}\int_{V':~t,\,t'=const}
\Theta^{\lambda\sigma}\Big(\mathbf{x}=\mathbf{x}(\mathbf{x}',t'),t\Big)\mathrm{d}^3x'.
\label{eq15}
\end{align} 
Note that $\mathbf{x}=\mathbf{x}(\mathbf{x}',t')$ in $\Theta^{\lambda\sigma}\big(\mathbf{x}=\mathbf{x}(\mathbf{x}',t'),t\big)$ denotes Eq.\,(\ref{eq3}).  By the change of variables $\mathbf{x}(\mathbf{x}',t')=\mathbf{x}$  or $(x',y',z';t')\rightarrow (x,y,z)$ with $t'$  as a constant parameter, from above Eq.\,(\ref{eq15}) we obtain 
\begin{equation}
P'^{\mu}=\frac{\partial X'^{\mu}}{\partial X^{\lambda}}\frac{\partial X'^{4}}{\partial X^{\sigma}}\frac{1}{\gamma}\int_{V:~t=const}
\Theta^{\lambda\sigma}(\mathbf{x},t)\mathrm{d}^3x,
\label{eq16}
\end{equation} 
where $\mathrm{d}^3x=|\partial(x,y,z)/\partial(x',y',z')|\mathrm{d}^3x'=\gamma\mathrm{d}^3x'$~ is employed, with the Jacobian determinant $\partial(x,y,z)/\partial(x',y',z')=\gamma$ being explained as the effect of Lorentz contraction physically (confer Appendix \ref{appa}).  
Since $t'$  is introduced as a constant parameter in the change of variables, $P'^{\mu}$  is independent of $t'$.

From Eq.\,(\ref{eq16}), with $\partial X'^{4}/\partial X^{4}=\gamma$ and the definition given by Eq.\,(\ref{eq7}), $P^{\mu}=\int_{V:~t=const}\Theta^{\mu 4}(\mathbf{x},t)\mathrm{d}^3x$, taken into account, we have 
\begin{align}
P'^{\mu}&=\frac{\partial X'^{\mu}}{\partial X^{\lambda}}\frac{\partial X'^{4}}{\partial X^{j}}\frac{1}{\gamma}\int_{V:~t=const}
\Theta^{\lambda j}(\mathbf{x},t)\mathrm{d}^3x
\notag \\
&\quad\quad~+~\frac{\partial X'^{\mu}}{\partial X^{\lambda}}\frac{\partial X'^{4}}{\partial X^{4}}\frac{1}{\gamma}\int_{V:~t=const}
\Theta^{\lambda 4}(\mathbf{x},t)\mathrm{d}^3x 
\notag \\
&=\frac{\partial X'^{\mu}}{\partial X^{\lambda}}\frac{\partial X'^{4}}{\partial X^{j}}\frac{1}{\gamma}\int_{V:~t=const}
\Theta^{\lambda j}(\mathbf{x},t)\mathrm{d}^3x
\notag \\
&\quad\quad~+~\frac{\partial X'^{\mu}}{\partial X^{\lambda}}\int_{V:~t=const}
\Theta^{\lambda 4}(\mathbf{x},t)\mathrm{d}^3x 
\notag  \\
&=\frac{\partial X'^{\mu}}{\partial X^{\lambda}}\frac{\partial X'^{4}}{\partial X^{j}}\frac{1}{\gamma}\int_{V:~t=const}
\Theta^{\lambda j}(\mathbf{x},t)\mathrm{d}^3x 
\notag \\
&\quad\quad~+~\frac{\partial X'^{\mu}}{\partial X^{\lambda}}P^{\lambda}, \quad (\mathrm{with}\quad j=1,2,3).
\label{eq17}
\end{align}  \\
If $P^{\mu}$  is a Lorentz four-vector, then 
\begin{equation}
P'^{\mu}=\frac{\partial X'^{\mu}}{\partial X^{\lambda}}P^{\lambda}
\label{eq18}
\end{equation} 
must hold.  Inserting Eq.\,(\ref{eq18}) into Eq.\,(\ref{eq17}), we have 
\begin{equation}
\frac{\partial X'^{\mu}}{\partial X^{\lambda}}\frac{\partial X'^{4}}{\partial X^{j}}\frac{1}{\gamma}\int_{V:~t=const}
\Theta^{\lambda j}(\mathbf{x},t)\mathrm{d}^3x=0,
\label{eq19}
\end{equation} \\
where $\big(\partial X'^{\mu}/\partial X^{\lambda}\big)$  is the Lorentz transformation matrix, with its determinant $\mathrm{det}\big(\partial X'^{\mu}/\partial X^{\lambda}\big)=1$ \cite[p.\,544]{r4}. With both sides of above Eq.\,(\ref{eq19}) multiplied by $\gamma$, from Eq.\,(\ref{eq1}) we have \\
\begin{align}
&\left( \begin{array}{cccc} 
1+\xi\beta_x^2 & \xi\beta_x\beta_y & \xi\beta_x\beta_z & -\gamma\beta_x \\ 
\xi\beta_y\beta_x & 1+\xi\beta_y^2 & \xi\beta_y\beta_z & -\gamma\beta_y \\ 
\xi\beta_z\beta_x & \xi\beta_z\beta_y & 1+\xi\beta_z^2  & -\gamma\beta_z \\ 
-\gamma\beta_x   & -\gamma\beta_y    & -\gamma\beta_z &  \gamma   \end{array} \right)
\notag \\
\notag \\
&\quad\times\left(\begin{array}{ccc}
a^{11} & a^{12} & a^{13} \\
a^{21} & a^{22} & a^{23} \\
a^{31} & a^{32} & a^{33} \\
a^{41} & a^{42} & a^{43} \end{array} \right)
\left(\begin{array}{c} -\gamma\beta_x \\ -\gamma\beta_y \\ -\gamma\beta_z \end{array} \right)
=
\left(\begin{array}{c} 0 \\ 0 \\ 0 \\ 0  \end{array} \right),
\label{eq20}
\end{align} 
where $a^{\lambda j}=\int_{V:~t=const}
\Theta^{\lambda j}(\mathbf{x},t)\mathrm{d}^3x$, with $\lambda=$1,2,3,4 and $j=$1,2,3. 

From above it is seen that Eq.\,(\ref{eq20})$\Leftrightarrow$Eq.\,(\ref{eq18}) through Eq.\,(\ref{eq17}) is valid.  Thus for $P^{\mu}$ to be a Lorentz four-vector, the sufficient and necessary condition is given by 
\begin{align}
a^{\lambda j}=&\int_{V:~t=const}
\Theta^{\lambda j}(\mathbf{x},t)\mathrm{d}^3x=0 
\notag \\
\notag \\
&\quad \mathrm{for}\quad\lambda=1,2,3,4 \quad\mathrm{and}\quad j=1,2,3.
\label{eq21}
\end{align} \\
The sufficiency of Eq.\,(\ref{eq21}) is apparent because we directly have Eq.\,(\ref{eq21})$\Rightarrow$Eq.\,(\ref{eq20})$\Rightarrow$Eq.\,(\ref{eq19})$\Rightarrow$Eq.\,(\ref{eq18}) from Eq.\,(\ref{eq17}).  The necessity is based on the fact that a four-vector must follow Lorentz rule between any two inertial frames, namely $\mb{\beta}c$ is arbitrary, and thus $a^{\lambda j}=0$  must hold for all $\lambda$  and $j$, because $(\beta_{x}\neq 0,\beta_{y}=0,\beta_{z}=0)\Rightarrow a^{\lambda 1}=0$, $(\beta_{x}=0,\beta_{y}\neq 0,\beta_{z}=0)\Rightarrow a^{\lambda 2}=0$, and $(\beta_{x}=0,\beta_{y}=0,\beta_{z}\neq 0)\Rightarrow a^{\lambda 3}=0$.  Thus we finish the proof of the sufficiency and necessity. \\

\noindent \textbf{Theorem 3.} Suppose that $\mathnormal{\Lambda}^{\mu}(\mathbf{x},t)=(\mb{\Lambda},\mathnormal{\Lambda}^{4})$ is an integrable Lorentz four-vector, defined in the domain $V$ in the laboratory frame $XYZ$, where $\mu$ = 1, 2, 3, and 4, with the index 4 corresponding to time component, and $V$ including its boundary is at rest in $XYZ$, namely any $\mathbf{x}\in V$ is independent of $t$.  The Lorentz scalar theorem states: The time-element space integral \\
\begin{equation}
\Phi=\int_{V:~t=const}\mathnormal{\Lambda}^{4}(\mathbf{x},t)\mathrm{d}^3x
\label{eq22}
\end{equation} \\
is a Lorentz scalar \emph{if and only if} \\
\begin{align}
&\int_{V:~t=const}\mathnormal{\Lambda}^{i}(\mathbf{x},t)\mathrm{d}^3x=0 \quad\mathrm{for}\quad i=1,2,3\quad\mathrm{or}\quad
\notag \\
\notag \\
&\int_{V:~t=const}\mb{\Lambda}(\mathbf{x},t)\mathrm{d}^3x=0
\label{eq23}
\end{align} \\
holds.\\

\noindent \textbf{Proof.} Corresponding to $\Phi=\int_{V:~t=const}\mathnormal{\Lambda}^{4}(\mathbf{x},t)\mathrm{d}^3x$ given by Eq.\,(\ref{eq22}), we first have to define $\Phi'=\int\mathnormal{\Lambda}'^{4}\mathrm{d}^3x'$ in $X'Y'Z'$, because the implication of $\Phi'=\int\mathnormal{\Lambda}'^{4}\mathrm{d}^3x'$  itself is ambiguous before the dependence of $\mathnormal{\Lambda}'^{4}$ on $\mathbf{x}',t',$ and $t$ is defined.  For this end, from Lorentz transformation we have
\begin{align}
\mathnormal{\Lambda}'^{4}(\mathbf{x},t)&=\frac{\partial X'^{4}}{\partial X^{\lambda}}\mathnormal{\Lambda}^{\lambda}(\mathbf{x},t)~~
\nonumber \\
\notag \\
\Rightarrow~\mathnormal{\Lambda}'^{4}\Big(\mathbf{x}=\mathbf{x}(\mathbf{x}',t'),t\Big)&=\frac{\partial X'^{4}}{\partial X^{\lambda}}\mathnormal{\Lambda}^{\lambda}\Big(\mathbf{x}=\mathbf{x}(\mathbf{x}',t'),t\Big)~~~~
\label{eq24}
\end{align}
where the space variables $\mathbf{x}$ in $\mathnormal{\Lambda}^{\lambda}(\mathbf{x},t)$ are replaced by $\mathbf{x}=\mathbf{x}(\mathbf{x}',t')$, namely the space Lorentz transformation Eq.\,(\ref{eq3}), but $t$  in $\mathnormal{\Lambda}^{\lambda}(\mathbf{x},t)$ is kept as it is.

Making integration in Eq.\,(\ref{eq24}) with respect to $(x, y, z)$ over $V$ in the laboratory frame, we have \\
\begin{align}
&\int_{V:~t=const}\mathnormal{\Lambda}'^{4}\Big(\mathbf{x}=\mathbf{x}(\mathbf{x}',t'),t\Big)\mathrm{d}^3x 
\notag \\
\notag \\
&\quad\quad=\frac{\partial X'^{4}}{\partial X^{\lambda}}\int_{V:~t=const}\mathnormal{\Lambda}^{\lambda}\Big(\mathbf{x}=\mathbf{x}(\mathbf{x}',t'),t\Big)\mathrm{d}^3x.
\label{eq25}
\end{align} \\
By the change of variables $(x,y,z)\rightarrow (x',y',z';t')$  with $t'$ as a constant parameter in the left-hand side of Eq.\,(\ref{eq25}), while keeping the integrals of the right-hand side to be computed in $XYZ$ frame, we obtain 
\begin{align}
&\int_{V':~t,\,t'=const}\mathnormal{\Lambda}'^{4}\Big(\mathbf{x}=\mathbf{x}(\mathbf{x}',t'),t\Big)\gamma\mathrm{d}^3x'
\notag \\ 
\notag \\
&\quad\quad=\frac{\partial X'^{4}}{\partial X^{\lambda}}\int_{V:~t=const}\mathnormal{\Lambda}^{\lambda}(\mathbf{x},t)\mathrm{d}^3x.
\label{eq26}
\end{align} 
where $\mathrm{d}^3x=|\partial(x,y,z)/\partial(x',y',z')|\mathrm{d}^3x'=\gamma\mathrm{d}^3x'$ is taken into account, with $\partial(x,y,z)/\partial(x',y',z')=\gamma$ the Jacobian determinant.

We define
\begin{equation}
\Phi'=\int_{V':~t',\,t=const}\mathnormal{\Lambda}'^{4}\Big(\mathbf{x}=\mathbf{x}(\mathbf{x}',t'),t\Big)\mathrm{d}^3x',
\label{eq27}
\end{equation} \\
where $\mathnormal{\Lambda}'^{4}\big(\mathbf{x}=\mathbf{x}(\mathbf{x}',t'),t\big)$ is defined in Eq.\,(\ref{eq24}).  Since $t'$  is introduced as a constant parameter in the change of variables in the space integral, $\Phi'$  does not contain $t'$  although the integrand $\mathnormal{\Lambda}'^{4}\big(\mathbf{x}=\mathbf{x}(\mathbf{x}',t'),t\big)$ in Eq.\,(\ref{eq27}) contains $t'$.  Thus with the both sides of Eq.\,(\ref{eq26}) divided by $\gamma$ and then Eq.\,(\ref{eq27}) inserted, we have 
\begin{align}
\Phi'&=\int_{V':~t',\,t=const}\mathnormal{\Lambda}'^{4}\Big(\mathbf{x}=\mathbf{x}(\mathbf{x}',t'),t\Big)\mathrm{d}^3x' 
\notag \\
& \notag \\
&=\frac{\partial X'^{4}}{\partial X^{\lambda}}\frac{1}{\gamma}\int_{V:~t=const}\mathnormal{\Lambda}^{\lambda}(\mathbf{x},t)\mathrm{d}^3x 
\notag \\
& \nonumber \\
&=\frac{\partial X'^{4}}{\partial X^{4}}\frac{1}{\gamma}\int_{V:~t=const}\mathnormal{\Lambda}^{4}(\mathbf{x},t)\mathrm{d}^3x
\notag \\ 
& \notag \\
&\quad\quad\quad+~\frac{\partial X'^{4}}{\partial X^{i}}\frac{1}{\gamma}\int_{V:~t=const}\mathnormal{\Lambda}^{i}(\mathbf{x},t)\mathrm{d}^3x 
\nonumber \\
&\hspace{22mm}(\mathrm{with}\quad i=1,2,3) 
\notag \\
& \notag \\
&=\Phi-\mb{\beta}\cdot\int_{V:~t=const}\mb{\Lambda}(\mathbf{x},t)\mathrm{d}^3x,
\label{eq28}
\end{align}  
where $\partial X'^{4}/\partial X^{4}=\gamma$, $(\partial X'^{4}/\partial X^{i})\mathnormal{\Lambda}^{i}=-\gamma\mb{\beta}\cdot\mb{\Lambda}$, and the definition $\Phi=\int_{V:~t=const}\mathnormal{\Lambda}^{4}(\mathbf{x},t)\mathrm{d}^3x$ given by Eq.\,(\ref{eq22}) are employed.

From Eq.\,(\ref{eq28}) we obtain the sufficient and necessary condition for $\Phi=\Phi'$  (Lorentz scalar), given by \\
\begin{align}
&\int_{V:~t=const}\mathnormal{\Lambda}^{i}(\mathbf{x},t)\mathrm{d}^3x=0 \quad\mathrm{for}\quad i=1,2,3\quad\mathrm{or}
\notag \\
\notag \\
&\int_{V:~t=const}\mb{\Lambda}(\mathbf{x},t)\mathrm{d}^3x=0.
\label{eq29}
\end{align} \\
The sufficiency is apparent, while the necessity comes from the fact that $\mb{\beta}$ is arbitrary.  Thus we complete the proof.\\

There are some main points to understand Theorem 3 that should be noted: \vspace{2mm} 

(i) If $\mathnormal{\Lambda}^{\mu}(\mathbf{x},t)$ is independent of $t$, namely $\partial\mathnormal{\Lambda}^{\mu}/\partial t \equiv 0$, then both $\Phi$  and $\Phi'$  are constants.\vspace{2mm} 

(ii) The divergence-less ($\partial_{\mu}\mathnormal{\Lambda}^{\mu}=0$) is not required, and there are no boundary conditions imposed on $\mathnormal{\Lambda}^{\mu}(\mathbf{x},t)$.\vspace{2mm} 

(iii) \emph{Asymmetry arising from resting $V$ and moving $V'$.}  Directly from Eq.\,(\ref{eq2}), we have \\
\begin{align*}
\mathnormal{\Lambda}'^{4}&=\gamma (\mathnormal{\Lambda}^{4}-\mb{\beta}\cdot\mb{\Lambda})~~
\notag \\
\notag \\
\Rightarrow~~\int_{V'}\mathnormal{\Lambda}'^{4}\mathrm{d}^3x'&=\int_{V'}\gamma (\mathnormal{\Lambda}^{4}-\mb{\beta}\cdot\mb{\Lambda})\mathrm{d}^3x'~~
\notag \\
\notag \\
\Rightarrow\hspace{16.4mm}\Phi'&=\Phi-\mb{\beta}\cdot\int_{V}\mb{\Lambda}\mathrm{d}^3x
\end{align*} \\
with $\mathrm{d}^3x'=\mathrm{d}^3x/\gamma$ used, namely Eq.\,(\ref{eq28}).  Conversely, from Eq.\,(\ref{eq4}) we have \\
\begin{align*}
\mathnormal{\Lambda}^{4}&=\gamma (\mathnormal{\Lambda}'^{4}-\mb{\beta}'\cdot\mb{\Lambda}')~~
\notag \\
\notag \\
\Rightarrow~~\int_{V}\mathnormal{\Lambda}^{4}\mathrm{d}^3x&=\int_{V}\gamma (\mathnormal{\Lambda}'^{4}-\mb{\beta}'\cdot\mb{\Lambda}')\mathrm{d}^3x~~
\notag \\
\notag \\
\Rightarrow\hspace{14.6mm}\Phi&=\gamma^2\Phi'-\gamma^2\mb{\beta}'\cdot\int_{V'}\mb{\Lambda}'\mathrm{d}^3x'
\end{align*} \\
with $\mathrm{d}^3x=\gamma\mathrm{d}^3x'$ used.  We find that \\
\begin{align*}
\Phi'&=\Phi-\mb{\beta}\cdot\int_{V}\mb{\Lambda}\mathrm{d}^3x ~~\quad\mathrm{and}\quad~~
\notag \\
\notag \\
\Phi&=\gamma^2\Phi'-\gamma^2\mb{\beta}'\cdot\int_{V'}\mb{\Lambda}'\mathrm{d}^3x'
\end{align*} 
are not symmetric, although 
\begin{align*}
\mathnormal{\Lambda}'^{4}&=\gamma (\mathnormal{\Lambda}^{4}-\mb{\beta}\cdot\mb{\Lambda}) ~~\quad\mathrm{and}\quad~~
\notag \\
\mathnormal{\Lambda}^{4}&=\gamma (\mathnormal{\Lambda}'^{4}-\mb{\beta}'\cdot\mb{\Lambda}')
\end{align*} 
are symmetric.  This asymmetry comes from the fact that $V$ is fixed in $XYZ$, while $V'$ is moving in $X'Y'Z'$. 

\section{Invalidity of M\o ller's theorem}
\label{s3}
In this section, (i) M\o ller's theorem is proved to be incorrect; (ii) based on Theorem 1, a counterexample of M\o ller's theorem is given; and (iii) a corrected version of M\o ller's theorem is provided, with a detailed elucidation given of why the corrected M\o ller's theorem only defines a trivial zero four-vector for EM stress--energy tensor. \\

\noindent \textbf{M\o ller's theorem.}  Suppose that $\Theta^{\mu\nu}(\mathbf{x},t)$ is an integrable Lorentz four-tensor, defined in the domain $V$ in the laboratory frame $XYZ$,  where $\mu, \nu$ = 1, 2, 3, and 4, with the index 4 corresponding to time component, and $V$ including its boundary is at rest in $XYZ$, namely any $\mathbf{x}\in V$ is independent of $t$.  All the elements of the tensor have first-order partial derivatives with respect to time-space coordinates  $X^{\mu}=(\mathbf{x},ct)$.  M\o ller's theorem states: If $\Theta^{\mu\nu}(\mathbf{x},t)$ is divergence-less ($\partial_{\nu}\Theta^{\mu\nu}(\mathbf{x},t)=0$), and $\Theta^{\mu\nu}(\mathbf{x},t)=0$  holds on the boundary of $V$ for any time ($-\infty<t<+\infty$) --- \emph{zero boundary condition}, then the time-column space integrals
\begin{equation}
P^{\mu}=\int_{V:~t=const}\Theta^{\mu 4}(\mathbf{x},t)\mathrm{d}^3x
\label{eq30}
\end{equation}
constitute a Lorentz four-vector \cite[pp.166-169]{r3}.\\ \\

\noindent \textbf{Proof.}  From M\o ller's sufficient condition, we first demonstrate that the time-column space integrals, given by Eq.\,(\ref{eq30}), are time-independent ($\partial P^{\mu}/\partial t\equiv 0$), then we prove that the sufficient condition is not enough to make Eq.\,(\ref{eq30}) be a four-vector, and we conclude that M\o ller's theorem is incorrect. \vspace{1mm}

Since  $\Theta^{\mu\nu}(\mathbf{x},t)=0$ holds on the boundary of $V$, using 3-dimensional Gauss's divergence theorem we obtain 
\begin{equation}
\int\limits_{V:~t=const}\partial_{i}\Theta^{\mu i}(\mathbf{x},t)\mathrm{d}^3x=0, \quad\mathrm{with}\quad i=1,2,3.
\label{eq31}
\end{equation}
Because the boundary of $V$ is at rest in the laboratory frame, we have 
\begin{equation}
\int\limits_{V:~t=const}\frac{\partial}{\partial t}(~\cdots~)\mathrm{d}^3x=\frac{\partial}{\partial t}\int\limits_{V:~t=const}(~\cdots~)\mathrm{d}^3x.
\label{eq32}
\end{equation}
From $\partial_{\nu}\Theta^{\mu\nu}(\mathbf{x},t)=0$, with Eq.\,(\ref{eq31}), Eq.\,(\ref{eq32}), and $X^{4}=ct$ taken into account, we have
\begin{align}
0&=\int\limits_{V:~t=const}\partial_{\nu}\Theta^{\mu\nu}(\mathbf{x},t)\mathrm{d}^3x  
\notag \\
&=\int\limits_{V:~t=const}\partial_{i}\Theta^{\mu i}(\mathbf{x},t)\mathrm{d}^3x~
\notag \\
&\hspace{15mm}+\quad \int\limits_{V:~t=const}\partial_{4}\Theta^{\mu 4}(\mathbf{x},t)\mathrm{d}^3x \notag \\
&=\int\limits_{V:~t=const}\partial_{4}\Theta^{\mu 4}(\mathbf{x},t)\mathrm{d}^3x 
\notag \\
&=\frac{\partial}{\partial (ct)}\int\limits_{V:~t=const}\Theta^{\mu 4}(\mathbf{x},t)\mathrm{d}^3x.
\label{eq33}
\end{align} \\
Inserting Eq.\,(\ref{eq30}) into above Eq.\,(\ref{eq33}) yields \\
\begin{equation}
\frac{\partial P^{\mu}}{\partial t}=\frac{\partial}{\partial t}\int\limits_{V:~t=const}\Theta^{\mu 4}(\mathbf{x},t)\mathrm{d}^3x\equiv 0.
\label{eq34}
\end{equation} 
Thus $P^{\mu}=\int_{V:~t=const}\Theta^{\mu 4}(\mathbf{x},t)\mathrm{d}^3x$ is constant although the integrand $\Theta^{\mu 4}(\mathbf{x},t)$ may depend on $t$.  However it should be emphasized that  
\begin{align}
&\frac{\partial}{\partial t}\int\limits_{V:~t=const}\Theta^{\mu j}(\mathbf{x},t)\mathrm{d}^3x=0 
\notag \\
&\hspace{20mm}\mathrm{for}\quad j=1,2,3 \quad\textrm{may not hold.}
\label{eq35}
\end{align} 

From the divergence-less ($\partial_{\nu}\Theta^{\mu\nu}=0$) and the zero-boundary condition ($\Theta^{\mu\nu}=0$ on boundary), we have achieved a conclusion that the time-column space integrals $P^{\mu}=\int_{V:~t=const}\Theta^{\mu 4}(\mathbf{x},t)\mathrm{d}^3x$  are time-independent constants.  In what follows, we will show that the divergence-less and the zero-boundary condition is not sufficient to make $P^{\mu}$  be a four-vector.  In other words, M\o ller's sufficient condition is not sufficient. 

From Eqs.\,(\ref{eq15})-(\ref{eq17}) in the proof of Theorem 1, we have 
\begin{align}
\label{eq36}
P'^{\mu}&=\int_{V':~t,\,t'=const}\Theta'^{\mu 4}\Big(\mathbf{x}=\mathbf{x}(\mathbf{x}',t'),t\Big)\mathrm{d}^3x'
\notag \\
& \notag \\
&=\frac{\partial X'^{\mu}}{\partial X^{\lambda}}\frac{\partial X'^{4}}{\partial X^{\sigma}}\frac{1}{\gamma}\int_{V:~t=const}
\Theta^{\lambda\sigma}(\mathbf{x},t)\mathrm{d}^3x 
\notag \\
& \notag \\
&=\frac{\partial X'^{\mu}}{\partial X^{\lambda}}\frac{\partial X'^{4}}{\partial X^{j}}\frac{1}{\gamma}\int_{V:~t=const}
\Theta^{\lambda j}(\mathbf{x},t)\mathrm{d}^3x 
\notag \\
&\hspace{10mm}\textrm{(allowed to be \emph{t}-dependent)} 
\notag \\
&\notag \\
&\hspace{10mm}+\quad\frac{\partial X'^{\mu}}{\partial X^{\lambda}}P^{\lambda}, \hspace{5mm} (\mathrm{with}\quad j=1,2,3). \\
&\hspace{12mm}\textrm{(\emph{t}-independent)}
\notag 
\end{align}  
Thus like Eq.\,(\ref{eq17}),  we obtain a \emph{sufficient and necessary} condition for constant $P^{\mu}$ to be a Lorentz four-vector, given below
\begin{align}
&a^{\lambda j}=\int_{V:~t=const}
\Theta^{\lambda j}(\mathbf{x},t)\mathrm{d}^3x=0 
\notag \\
&\hspace{27mm}\mathrm{for}~\lambda=1,2,3,4 ~\mathrm{and}~ j=1,2,3,
\label{eq37}
\end{align}
which is the same as Eq.\,(\ref{eq10}). However M\o ller's sufficient condition does not include this \emph{sufficient and necessary} condition, and accordingly, M\o ller's theorem is fundamentally wrong.  Thus we finish the proof. \\

\noindent \emph{Counterexample of M\o ller's theorem.}  To further convince readers, given below is a pure mathematical counterexample to disprove M\o ller's theorem based on Theorem 1.  As indicated in Sec.\,\ref{s5} later, this counterexample of M\o ller's theorem is also the counterexample of Landau-Lifshitz and Weinberg's versions of Laue's theorem \cite{r6}.

Suppose that there is a symmetric Lorentz four-tensor 
\begin{equation}
A^{\mu\nu}=\left( \begin{array}{cccc} 
0 & 0 & 0 & f(\mathbf{x}) \\ 
0 & 0 & 0 & f(\mathbf{x}) \\ 
0 & 0 & 0 & f(\mathbf{x}) \\ 
f(\mathbf{x})   & f(\mathbf{x})    & f(\mathbf{x}) &  -(ct)(f_x+f_y+f_z)   \end{array} \right),
\label{eq38}
\end{equation} \\
defined in the cubic domain $V ~(-\pi\le x,y,z\le \pi)$, where $f(\mathbf{x})=(\sin x)^2(\sin y)^2(\sin z)^2$  is independent of time, with $\int_{V}f(\mathbf{x})\mathrm{d}^3x=\pi^3$, and  $f_x\equiv \partial f/\partial x$,  $f_y\equiv \partial f/\partial y$, and  $f_z\equiv \partial f/\partial z$.  $A^{\mu\nu}$ is divergence-less ($\partial_{\mu}A^{\mu\nu}=0\Leftrightarrow\partial_{\nu}A^{\mu\nu}=0$  because of  $A^{\mu\nu}=A^{\nu\mu}$), and satisfies the M\o ller's zero boundary condition: $A^{\mu\nu}$=0  holds on the boundary $x,y,z=\pm\pi$ for $-\infty<t<+\infty$. Thus   $A^{\mu\nu}$ satisfies the sufficient condition of M\o ller's theorem, and 
\begin{equation}
M^{\mu}=\int_{V:~t=const}A^{\mu 4}\mathrm{d}^3x=(\pi^3,\pi^3,\pi^3,0)
\label{eq39}
\end{equation} \\
is supposed to be a Lorentz four-vector.  

However because 
\begin{align}
\int_{V:~t=const}A^{41}\mathrm{d}^3x&=\int_{V:~t=const}A^{42}\mathrm{d}^3x 
\notag \\
\notag \\
&=\int_{V:~t=const}A^{43}\mathrm{d}^3x 
\notag \\
\notag \\
&=\pi^3\ne 0,
\label{eq40}
\end{align} 
$A^{\mu\nu}$ does not satisfy the sufficient and necessary condition Eq.\,(\ref{eq10}) of Theorem 1, and accordingly,   $M^{\mu}=\int_{V:~t=const}A^{\mu 4}\mathrm{d}^3x$ is not a four-vector.  Thus M\o ller's theorem is disproved by this counterexample based on Theorem 1.\\

The above counterexample shows that the sufficient condition of M\o ller's theorem indeed does not includes the sufficient and necessary condition Eq.\,(\ref{eq10}) of Theorem 1. Obviously, M\o ller's theorem can be easily corrected by adding the condition Eq.\,(\ref{eq10}), as follows.\\

\noindent \textbf{Corrected M\o ller's theorem.}  Suppose that $\Theta^{\mu\nu}(\mathbf{x},t)$ is an integrable Lorentz four-tensor, defined in the domain $V$ in the laboratory frame $XYZ$,  where $\mu, \nu$ = 1, 2, 3, and 4, with the index 4 corresponding to time component, and $V$ including its boundary is at rest in $XYZ$, namely any $\mathbf{x}\in V$ is independent of $t$.  It is assumed that  $\Theta^{\mu\nu}(\mathbf{x},t)$ is divergence-less ($\partial_{\nu}\Theta^{\mu\nu}(\mathbf{x},t)=0$), and $\Theta^{\mu\nu}(\mathbf{x},t)=0$ holds on the boundary of $V$ for any time ($-\infty<t<+\infty$) --- \emph{zero boundary condition}.  The corrected M\o ller's theorem states: The time-column space integrals 
\begin{equation}
P^{\mu}=\int_{V:~t=const}\Theta^{\mu 4}(\mathbf{x},t)\mathrm{d}^3x
\label{eq41}
\end{equation}
constitute a Lorentz four-vector \emph{if and only if}
\begin{align}
&\int_{V:~t=const}\Theta^{\mu j}(\mathbf{x},t)\mathrm{d}^3x=0
\notag \\ 
&\hspace{18mm}\mathrm{for}\quad\mu=1,2,3,4 \quad\mathrm{and}\quad j=1,2,3.
\label{eq42}
\end{align}
holds.\\

However we would like to indicate, by enumerating specific examples as follows, that the corrected M\o ller's theorem has a limited application. \\

\noindent \emph{Example 1 for corrected M\o ller's theorem.} Consider Minkowski EM stress--energy tensor for ``a pure radiation field in matter'' \cite{r10}, given by 
\begin{equation}
\widetilde{T}^{\mu\nu}=(T^{\mu\nu})^{T}, ~~~~\mathrm{with}~~~~ T^{\mu\nu}=\left(\begin{array}{cc} \check{\mathbf{T}}_{\mathrm{M}} & c\mathbf{g}_{\mathrm{A}} \\ c\mathbf{g}_{\mathrm{M}} &W_{\mathrm{em}}\end{array}\right), 
\label{eq43}
\end{equation} 
where $\widetilde{T}^{\mu\nu}$ is the transpose of $T^{\mu\nu}$, with $\partial_{\nu}\widetilde{T}^{\mu\nu}=\partial_{\nu}T^{\nu\mu}=\big(\nabla\cdot\check{\mathbf{T}}_{\mathrm{M}}+\partial\mathbf{g}_{\mathrm{M}}/\partial t,~\nabla\cdot(c\mathbf{g}_{\mathrm{A}})+\partial W_{\mathrm{em}}/\partial (ct)\big)$; ~$\mathbf{g}_{\mathrm{A}}=\mathbf{E}\times\mathbf{H}/c^2$ is the Abraham momentum;  $\mathbf{g}_{\mathrm{M}}=\mathbf{D}\times\mathbf{B}$ is the Minkowski momentum; $W_{\mathrm{em}}=0.5(\mathbf{D}\cdot\mathbf{E}+\mathbf{B}\cdot\mathbf{H})$ is the EM energy density; and $\check{\mathbf{T}}_{\mathrm{M}}=-\mathbf{DE}-\mathbf{BH}+\check{\mathbf{I}}0.5(\mathbf{D}\cdot\mathbf{E}+\mathbf{B}\cdot\mathbf{H})$ is the Minkowski stress tensor, with $\check{\mathbf{I}}$ the unit tensor \cite{r6}.  We first assume that the corrected M\o ller's theorem is applicable for this EM tensor.  Then let us see what conclusion we can get.

The pre-assumption of corrected M\o ller's theorem is the tensor's divergence-less plus a zero-boundary condition.  The zero-boundary condition requires that all the tensor elements be equal to zero on the boundary for any time ($-\infty<t<+\infty$).  Thus for the EM stress--energy tensor given by Eq.\,(\ref{eq43}), the pre-assumption requires $\partial_{\nu}\widetilde{T}^{\mu\nu}=0$ holding within the finite domain $V$ of a physical system, and Poynting vector $\mathbf{E}\times\mathbf{H}=0$ and Minkowski momentum $\mathbf{D}\times\mathbf{B}=0$ holding on the boundary of $V$ for any time ($-\infty<t<+\infty$).

Physically, the pre-assumption is extremely strong and severe, because it requires that (i) within the domain $V$, there are no any sources ($\partial_{\nu}\widetilde{T}^{\mu\nu}=0$), and (ii) the EM energy and Minkowski momentum never flow through the closed boundary of $V$ for any time ($\mathbf{E}\times\mathbf{H}=0$ and $\mathbf{D}\times\mathbf{B}=0$ for $-\infty<t<+\infty$).  Thus this physical system is never provided with any EM energy and momentum.  According to energy--momentum conservation law, no EM fields can be supported within the domain $V$ in such a case, leading to a zero field solution.  Thus the corrected M\o ller's theorem only defines a trivial \emph{zero} four-vector for an EM stress--energy tensor of a \emph{finite closed} physical system, even if this theorem is applicable. \\

\noindent \emph{Example 2 for corrected M\o ller's theorem.}  Nevertheless, the corrected M\o ller's theorem may define a \emph{non-zero} four-vector in general.  As an example, consider the tensor given by 
\begin{equation}
B^{\mu\nu}(\mathbf{x},t)=\left( \begin{array}{cccc} 
0 & 0 & 0 & 0 \\ 
0 & 0 & 0 & 0 \\ 
0 & 0 & 0 & 0 \\ 
0 & 0 & 0 & f(\mathbf{x})\end{array} \right),
\label{eq44}
\end{equation}
defined in the cubic domain $V ~(-\pi\le x,y,z\le \pi)$, where  $f(\mathbf{x})=(\sin x)^2(\sin y)^2(\sin z)^2$, with  $\int_{V}f(\mathbf{x})\mathrm{d}^3x=\pi^3$.  $B^{\mu\nu}(\mathbf{x},t)$ is divergence-less ($\partial_{\nu}B^{\mu\nu}=0$), and satisfies the zero boundary condition: $B^{\mu\nu}(\mathbf{x},t)=0$  on the boundary $(x,y,z=\pm\pi)$ for $-\infty<t<+\infty$; thus the pre-assumption of corrected M\o ller's theorem is satisfied.  On the other hand, $\int_{V:~t=const}B^{\mu j}(\mathbf{x},t)\mathrm{d}^3x=0 $ holds for $\mu=1,2,3,4$ and $j=1,2,3$; thus $B^{\mu\nu}(\mathbf{x},t)$ also satisfies the sufficient and necessary condition Eq.\,(\ref{eq42}) for the corrected M\o ller's theorem.  Accordingly, $\int_{V:~t=const}B^{\mu 4}(\mathbf{x},t)\mathrm{d}^3x=(0,0,0,\pi^3)\ne 0$ is a four-vector --- the corrected M\o ller's theorem may define a \emph{non-zero} four-vector in general.\\

\noindent \emph{Conclusion for corrected M\o ller's theorem.}  In conclusion, the corrected M\o ller's theorem may define a non-zero four-vector in general; however, it only defines a trivial zero four-vector for an EM stress--energy tensor of a finite closed physical system.  Thus the application of the theorem is limited. \\

\noindent \emph{Differences between three four-vector theorems.}  We have three four-vector theorems: Theorem 1 and corrected M\o ller's theorem (both presented in the present paper), and generalized von Laue's theorem (presented in Ref.\,\cite{r6}).  For the convenience to compare, we write down the generalized von Laue's theorem from Ref.\,\cite{r6} as follows.\\

\noindent \textbf{Generalized von Laue's theorem.}  Assume that $\Theta^{\mu\nu}(\mathbf{x})$ is an integrable Lorentz four-tensor, defined in the domain $V$ in the laboratory frame $XYZ$, where $\mu, \nu$ = 1, 2, 3, and 4, with the index 4 corresponding to time component, $V$ including its boundary is at rest in $XYZ$, and $\Theta^{\mu\nu}$  is independent of time ($\partial\Theta^{\mu\nu}/\partial t\equiv 0$).  The generalized von Laue's theorem states: The time-column-element space integrals $P^{\mu}=\int_{V}\Theta^{\mu 4}(\mathbf{x})\mathrm{d}^3x$ constitute a Lorentz four-vector \emph{if and only if} $\int_{V}\Theta^{\mu j}(\mathbf{x})\mathrm{d}^3x=0$ holds for all $\mu$ =1, 2, 3, 4  and  $j$ = 1, 2, 3.\\

Between the corrected M\o ller's theorem and the above generalized von Laue's theorem, the difference is that in the corrected M\o ller's theorem, the divergence-less ($\partial_{\nu}\Theta^{\mu\nu}=0$) plus a zero boundary condition ($\Theta^{\mu\nu}=0$ on boundary) is taken as a pre-assumption, and $\Theta^{\mu\nu}(\mathbf{x},t)$ is allowed to be time-dependent, while in the generalized von Laue's theorem, $\partial\Theta^{\mu\nu}/\partial t\equiv 0$ is taken as a pre-assumption, and $\Theta^{\mu\nu}(\mathbf{x},t)\equiv\Theta^{\mu\nu}(\mathbf{x})$  is not allowed to be time-dependent, but no boundary condition is required.  Compared with the corrected M\o ller's theorem and the generalized von Laue's theorem, Theorem 1 does not have any pre-assumption; however, the three theorems have the same definition $P'^{\mu}$, as shown below. 

From Eq.\,(\ref{eq36}), we know that the same definition of $P'^{\mu}$ is used in both Theorem 1 and the corrected M\o ller's theorem, given by
\begin{align}
P'^{\mu}&=\int_{V':~t,\,t'=const}\Theta'^{\mu 4}\Big(\mathbf{x}=\mathbf{x}(\mathbf{x}',t'),t\Big)\mathrm{d}^3x' 
\notag \\
\notag \\
&=\frac{\partial X'^{\mu}}{\partial X^{\lambda}}\frac{\partial X'^{4}}{\partial X^{\sigma}}\frac{1}{\gamma}\int_{V:~t=const}
\Theta^{\lambda\sigma}(\mathbf{x},t)\mathrm{d}^3x. 
\label{eq45}
\end{align}  \\
If $\Theta^{\mu\nu}(\mathbf{x},t)$ is independent of $t$, namely   $\Theta^{\mu\nu}(\mathbf{x},t)\equiv\Theta^{\mu\nu}(\mathbf{x})$, then the above Eq.\,(\ref{eq45}) becomes \\
\begin{align}
P'^{\mu}&=\int_{V':~t,\,t'=const}\Theta'^{\mu 4}\Big(\mathbf{x}=\mathbf{x}(\mathbf{x}',t')\Big)\mathrm{d}^3x'
\notag \\
\notag \\
&=\frac{\partial X'^{\mu}}{\partial X^{\lambda}}\frac{\partial X'^{4}}{\partial X^{\sigma}}\frac{1}{\gamma}\int_{V:~t=const}
\Theta^{\lambda\sigma}(\mathbf{x})\mathrm{d}^3x. 
\label{eq46}
\end{align} \\
This is exactly the case of von Laue's theorem presented in Ref.\,\cite{r6}, where $\Theta'^{\mu 4}\big(\mathbf{x}=\mathbf{x}(\mathbf{x}',t')\big)$ is written as  $\Theta'^{\mu 4}(\mathbf{x}',ct')$, and $t$  does not show up. \\

\noindent \emph{Adaptability of Theorem 1.}  Since Theorem 1 does not have a pre-assumption, it may have a better adaptability.  To show this, a specific example is given below.

Suppose that there is a symmetric Lorentz four-tensor \\
\begin{equation}
R^{\mu\nu}(\mathbf{x},t)=\left( \begin{array}{cccc} 
0 & 0 & 0 & x \\ 
0 & 0 & 0 & 0 \\ 
0 & 0 & 0 & 0 \\ 
x & 0 & 0 & -ct\end{array} \right),
\label{eq47}
\end{equation} \\
defined in the cubic domain $V~(-\pi\le x,y,z\le\pi)$ with its boundary S ($x,y,z=\pm\pi$). \vspace{2mm}

From Eq.\,(\ref{eq47}) we know that \vspace{2mm} \\
\indent (i) $\partial_{\nu}R^{\mu\nu}=0$ holds but $R^{\mu\nu}(\mathbf{x},t)$  does not have a zero-boundary condition \big($R^{41}(\mathbf{x},t)=\pi\ne 0$  on the boundary:  $x=\pi$ and $-\pi\le y,z\le\pi$, for example\big).  Thus the corrected M\o ller's theorem does not apply. \vspace{1mm} \\
\indent (ii)  $\partial R^{\mu\nu}/\partial t\equiv 0$  does not hold, because of  $\partial R^{44}/\partial t=-c\ne 0$ .  Thus the generalized von Laue's theorem does not apply either. \vspace{1mm} \\
\indent (iii) $\int_{V:~t=const}R^{\mu j}(\mathbf{x},t)\mathrm{d}^3x=0 $ for $\mu=1,2,3,4$ and $j=1,2,3$ holds, satisfying the sufficient and necessary condition Eq.\,(\ref{eq10}) of Theorem 1.  Thus $N^{\mu}=\int_{V:~t=const}R^{\mu 4}(\mathbf{x},t)\mathrm{d}^3x=\int_{V:~t=const}(x,0,0,-ct)\mathrm{d}^3x=(0,0,0,-8\pi^3ct)$ is a Lorentz four-vector. \vspace{2mm}

From above example we see that Theorem 1 has a better adaptability.  It is interesting to indicate that Theorem 1 can be used to analyze the EM stress--energy tensor for a plane light wave in a dielectric medium, as shown in Appendix \ref{appb}.

\section{Invalidity of Weinberg's claim}
\label{s4}
In relativistic electrodynamics, there are two main-stream arguments for the Lorentz invariance of total electric charge.  One of them comes from an assumption that the total electric charge is an experimental invariant, as presented in the textbook by Jackson \cite[p.555]{r4}; the other comes from a well-accepted ``invariant conservation law'' that the divergence-less of current density four-vector makes the total charge be a Lorentz scalar, as claimed in the textbook by Weinberg \cite[p.41]{r2}.  In this section, by enumerating a counterexample we use Theorem 3 to disprove Weinberg's claim. 

For a physical system defined in the domain $V$ with $S$ as its closed boundary, we will show that the divergence-less ($\partial_{\mu}J^{\mu}=0$) of current density four-vector $J^{\mu}=(\mathbf{J},c\rho)$ plus a boundary zero-integral given by $\oint_{~S}\mathbf{J}(\mathbf{x},t)\cdot\mathrm{d}\mathbf{S}=0$ makes the total charge $Q$ in $V$ be a time-independent constant; however, it is not enough to make the constant be a Lorentz scalar. \\

\noindent \emph{Constant of total electric charge.}  From $\partial_{\mu}J^{\mu}=0\Rightarrow\nabla\cdot\mathbf{J}+\partial\rho/\partial t=0$, with $Q=\int_{V}\rho\mathrm{d}^3x$ taken into account \cite[p.\,461]{r21}, we have $\oint_{~S}\mathbf{J}(\mathbf{x},t)\cdot\mathrm{d}\mathbf{S}+\mathrm{d}Q/\mathrm{d}t=0$, where the exchangeability between $\partial/\partial t$  and $\int_V$  is employed so that $\int_{V}(\partial\rho/\partial t)\mathrm{d}^3x=\mathrm{d}Q/\mathrm{d}t$ holds, as usually presented in textbooks \cite[p.\,20]{r2}\,\cite[p.\,345]{r41}.  If $\oint_{~S}\mathbf{J}(\mathbf{x},t)\cdot\mathrm{d}\mathbf{S}=0$ holds, then we have $\mathrm{d}Q/\mathrm{d}t=0\Rightarrow Q=const$ .  Physically, the current density $\mathbf{J}=\rho\mathbf{u}$  is a \emph{charge density flow}, where $\mathbf{u}$  is the charge moving velocity, and $\oint_{~S}\mathbf{J}(\mathbf{x},t)\cdot\mathrm{d}\mathbf{S}=0$ means that there is no net charge flowing into or out of $V$.  Thus the total electric charge $Q$ in $V$ is constant; however, it never means that this constant is a Lorentz invariant. (It is should be emphasized that only from $\partial_{\mu}J^{\mu}=0$ without $\oint_{~S}\mathbf{J}(\mathbf{x},t)\cdot\mathrm{d}\mathbf{S}=0$ considered, one cannot derive  $Q=const$  in $V$.) 

Why is the constant total charge $Q$ in $V$ in above, resulting from $\partial_{\mu}J^{\mu}=0\Rightarrow\nabla\cdot\mathbf{J}=-\partial\rho/\partial t$, not a Lorentz invariant?  That is because  $Q$ in $V$   in frame $XYZ$  and $Q'$ in $V'$ in frame $X'Y'Z'$  do not refer to the same total charge in the same volume physically, where $X'Y'Z'$  moves at $\mb{\beta}c\ne 0$  relatively to  $XYZ$.  This conclusion directly comes from the fact that, the volume $V$  in $\int_{V}(\partial\rho/\partial t)\mathrm{d}^3x$  is fixed in  $XYZ$, and the volume $V'$  in $\int_{V'}(\partial\rho'/\partial t')\mathrm{d}^3x'$  is fixed in  $X'Y'Z'$ so that $\partial/\partial t$  and  $\int_{V}$, and $\partial/\partial t'$  and  $\int_{V'}$ are exchangeable, respectively, as shown in Eq.\,(\ref{eq32}), in order to make both $\int_{V}(\partial\rho/\partial t)\mathrm{d}^3x=\mathrm{d}Q/\mathrm{d}t$  and $\int_{V'}(\partial\rho'/\partial t')\mathrm{d}^3x'=\mathrm{d}Q'/\mathrm{d}t'$ hold in general.  However according to Einstein's relativity \cite{r20}, $V$  fixed in $XYZ$  is moving observed in  $X'Y'Z'$.  Now $V'$  in $\int_{V'}(\partial\rho'/\partial t')\mathrm{d}^3x'$  is at rest in  $X'Y'Z'$, and thus $V$  and   $V'$ do not denote the same volume, and  $Q$  and  $Q'$  do not denote the same total charge. Therefore, the continuity equation $\partial_{\mu}J^{\mu}=0$  cannot be taken as the ``invariant [charge] conservation law''  in relativity \cite[p.\,40]{r2}, which can be better understood from the mathematical counterexample given below.\\

\noindent \emph{Counterexample of Weinberg's claim.}  Why is $\partial_{\mu}J^{\mu}=0$  not a sufficient condition to make  $Q=const$  be a Lorentz scalar, even additionally plus a zero-boundary condition  $J^{\mu}=(\mathbf{J},c\rho)=0$ on $S \Rightarrow\oint_{~S}\mathbf{J}\cdot\mathrm{d}\mathbf{S}=0$?  To understand this, consider a mathematical four-vector, given by 
\begin{align}
W^{\mu}(\mathbf{x}.t)=(\mathbf{W},W^4)
=\Big( f(\mathbf{x}),0,0,-(ct)f_x\Big)
\label{eq48} 
\end{align} 
defined in the cubic domain $V$ ($-\pi\le x,y,z\le\pi$), where $\mathbf{W}=\big(f(\mathbf{x}),0,0\big)$,  $W^4=-(ct)f_x$, and $f(\mathbf{x})=(\sin x)^2(\sin y)^2(\sin z)^2$, with $\int_{V}f(\mathbf{x})\mathrm{d}^3x=\pi^3$. $W^{\mu}(\mathbf{x},t)$ satisfies the zero-boundary condition, namely $W^{\mu}=(\mathbf{W},W^4)=0$ holds on the boundary $S$ ($x,y,z=\pm \pi$).

$W^{\mu}$ is divergence-less, namely $\partial_{\mu}W^{\mu}=0$, and in addition, $W^{\mu}$  has a zero-boundary condition $\Rightarrow\oint_{~S}\mathbf{W}\cdot\mathrm{d}\mathbf{S}=0$ holds.  According to Weinberg's claim, $\partial_{\mu}W^{\mu}=0$  makes $\int W^4\mathrm{d}^3x$ be a Lorentz scalar.

However because of $\int_{V:~t=const}\mathbf{W}\mathrm{d}^3x=(\pi^3,0,0)\ne 0$,  $W^{\mu}$ does not satisfy the sufficient and necessary condition Eq.\,(\ref{eq23}) of Theorem 3.  Thus according to Theorem 3,   $\Phi=\int_{V:~t=const}W^4\mathrm{d}^3x=\int_{V:~t=const}-(ct)f_x\mathrm{d}^3x~(=0)$ is not a Lorentz scalar.  To better understand this, from Eq.\,(\ref{eq28}) we have 
\begin{align}
\Phi'&=\Phi-\mb{\beta}\cdot\int_{V:~t=const}\mathbf{W}\mathrm{d}^3x
\notag \\
&=\Phi-\beta_x\int_{V}f(\mathbf{x})\mathrm{d}^3x 
\notag \\ 
&=\Phi-\beta_x\pi^3,
\label{eq49}
\end{align} \\
and $\Phi'=\Phi$ cannot hold for any  $\beta_x\ne 0$.  Thus Weinberg's claim is disproved, namely $\partial_{\mu}W^{\mu}=0$ is not a sufficient condition to make $\int W^4\mathrm{d}^3x$ be a scalar. 

In the above counterexample,  $\partial_{\mu}W^{\mu}=0$ holds but $\int \mathbf{W}\mathrm{d}^3x=0$ does not hold.  Thus in general, the sufficient condition  $\partial_{\mu}\mathnormal{\Lambda}^{\mu}=0$ of Weinberg's claim does not includes the sufficient and necessary condition $\int\mb{\Lambda}\mathrm{d}^3x=0$ of Theorem 3.

The current density four-vector $J^{\mu}=(\mathbf{J},c\rho)$  and the above counterexample $W^{\mu}=(\mathbf{W},W^4)$ are all divergence-less, while the Lorentz property of their time-element space integrals does not depends on the divergence-less.  Now let us take a look of four-vectors that are not divergence-less, and see what difference they may have.\vspace{2mm}

\noindent \emph{Example 1.}  Consider a four-vector, given by $\mathnormal{\Gamma}^{\mu}=(\mathbf{\Gamma},\mathnormal{\Gamma}^4)=(\sin x\sin y\sin z,0,0,0)$  defined in the cubic domain $V~(-\pi\le x,y,z\le\pi)$, with $\partial_{\mu}\mathnormal{\Gamma}^{\mu}=\cos x\sin y\sin z\ne 0$  holding except for some individual discrete points, and $\int_{V:~t=const}\mathbf{\Gamma}\mathrm{d}^3x=(0,0,0)=0$ holding.  According to Theorem 3, $\Phi=\int_{V:~t=const}\mathnormal{\Gamma}^4\mathrm{d}^3x~(=0)$ is a Lorentz scalar, because $\Phi'=\Phi-\mb{\beta}\cdot\int_{V:~t=const}\mathbf{\Gamma}\mathrm{d}^3x=\Phi$.  Put it simply, for  $\partial_{\mu}\mathnormal{\Gamma}^{\mu}\ne 0$, $\int\mathnormal{\Gamma}^4\mathrm{d}^3x$ is a Lorentz scalar. \vspace{2mm}

\noindent \emph{Example 2.}  Consider a four-vector, given by $U^{\mu}=(\mathbf{U},U^4)=(\sin^2 x\sin^2 y\sin^2 z,0,0,0)$  defined in the cubic domain  $V~(-\pi\le x,y,z\le\pi)$, with  $\partial_{\mu}U^{\mu}=2\cos x\sin x\sin^2 y\sin^2 z\ne 0$  holding except for some individual discrete points, and $\int_{V:~t=const}\mathbf{U}\mathrm{d}^3x=(\pi^3,0,0)\ne 0$ holding.  According to Theorem 3,     $\Phi=\int_{V:~t=const}U^4\mathrm{d}^3x~(=0)$ is not a Lorentz scalar, because $\Phi'=\Phi-\mb{\beta}\cdot\int_{V:~t=const}\mathbf{U}\mathrm{d}^3x=\Phi-\beta_x\pi^3\Rightarrow\Phi'\ne\Phi$ for any $\beta_x\ne 0$.  Put it simply, for $\partial_{\mu}U^{\mu}\ne 0$, $\int U^4\mathrm{d}^3x$ is not a Lorentz scalar. \vspace{2mm}

From above \emph{Example 1} and \emph{Example 2}, we know that $\int\mathnormal{\Gamma}^4\mathrm{d}^3x$  is a Lorentz scalar for $\partial_{\mu}\mathnormal{\Gamma}^{\mu}\ne 0$, and  $\int U^4\mathrm{d}^3x$ is not a Lorentz scalar for $\partial_{\mu}U^{\mu}\ne 0$.  We have known that $\int J^4\mathrm{d}^3x$  is a Lorentz scalar for $\partial_{\mu}J^{\mu}=0$ \cite{r6}, while the counterexample of Weinberg's claim tells us that  $\int W^4\mathrm{d}^3x$  is not a Lorentz scalar for  $\partial_{\mu}W^{\mu}=0$.  Thus we can generally conclude that whether the time-element space integral of a four-vector is a Lorentz scalar has nothing to do with the divergence-less property of the four-vector.

The invariance problem of total electric charge has been resolved by using ``derivative von Laue's theorem'' in Ref.\,\cite{r6}, which indicates that the invariance comes from two facts: (a) $J^{\mu}$ is a four-vector and (b) the moving velocity of any charged particles is less than vacuum light speed.  This strict theoretical result removes the assumption that the total charge is an experimental invariant \cite[p.555]{r4}.

The difference between the derivative von Laue's theorem \cite{r6} and Theorem 3 is that the derivative von Laue's theorem has a pre-assumption of $\partial\mathnormal{\Lambda}^{\mu}/\partial t\equiv 0$, namely $\mathnormal{\Lambda}^{\mu}=(\mathbf{\Lambda},\mathnormal{\Lambda}^4)$ is not allowed to be time-dependent, while Theorem 3 does not.  For example, we also can use the derivative von Laue's theorem \cite{r6} to analyze the four-vector $\mathnormal{\Gamma}^{\mu}=(\mathbf{\Gamma},\mathnormal{\Gamma}^4)=(\sin x\sin y\sin z,0,0,0)$ discussed above because $\partial\mathnormal{\Gamma}^{\mu}/\partial t\equiv 0$ holds, but we cannot use it to analyze $W^{\mu}(\mathbf{x},t)$ given by Eq.\,(\ref{eq48}), because $\partial W^{\mu}/\partial t\equiv 0$  does not hold.  Thus Theorem 3 has a better adaptability.

\section{Remarks and conclusions}
\label{s5}
In this paper, we have developed Lorentz four-vector theorems (Theorem 1 for \emph{column} four-vector and Theorem 2 for \emph{row} four-vector; they are essentially the same) and Lorentz scalar theorem (Theorem 3).  Based on Theorem 1, we find that the well-established M\o ller's theorem is fundamentally wrong, and we provided a corrected version of M\o ller's theorem (see Sec.\,\ref{s3}).  Based on Theorem 3, we disproved Weinberg's claim, and obtained a general conclusion for the Lorentz property of a four-vector's time-element space integral (see Sec.\,\ref{s4}). 

We have shown that the sufficient condition of M\o ller's theorem makes the time-column space integrals of a tensor be time-independent constants; however, it is not a sufficient condition to make the integrals constitute a Lorentz four-vector.  The corrected M\o ller's theorem has a limited application; especially for Minkowski EM stress--energy tensor, the corrected M\o ller's theorem only defines a trivial zero four-vector. 

We have shown that there are three four-vector theorems: (a) generalized von Laue's theorem; (b) corrected M\o ller's theorem; and (c) Theorem 1. The generalized von Laue's theorem, presented in Ref.\,\cite{r6}, has a pre-assumption that tensor $\Theta^{\mu\nu}$ is required to be time-independent $(\partial\Theta^{\mu\nu}/\partial t\equiv 0)$.  The corrected M\o ller's theorem, provided in the present paper, also has a pre-assumption that tensor $\Theta^{\mu\nu}$  is required to be divergence-less  $(\partial_{\nu}\Theta^{\mu\nu}=0)$ and required to satisfy a zero boundary condition ($\Theta^{\mu\nu}=0$ on boundary) but $\Theta^{\mu\nu}$ is allowed to be time-dependent.  Compared with the generalized von Laue's theorem and corrected M\o ller's theorem, Theorem 1 does not have any pre-assumption, while the three theorems have the same sufficient and necessary condition.  Thus Theorem 1 has a better adaptability, as shown by a specific example described by Eq.\,(\ref{eq47}) in Sec.\,\ref{s3}. 

However it should be noted that, just because the generalized von Laue's theorem has a pre-assumption of $\partial\Theta^{\mu\nu}(\mathbf{x},t)/\partial t\equiv 0$ (but no boundary condition required) and the corrected M\o ller's theorem has a pre-assumption of $\partial_{\nu}\Theta^{\mu\nu}(\mathbf{x},t)=0$ plus $\Theta(\mathbf{x},t)=0$ on boundary (zero boundary condition), the four-vector $P^{\mu}=\int_{V:~t=const}\Theta^{\mu 4}(\mathbf{x},t)\mathrm{d}^3x$ defined by the two theorems is time-independent ($\partial P^{\mu}/\partial t\equiv 0$).  Thus the generalized von Laue's theorem and the corrected M\o ller's theorem can be taken as ``conservation laws'' in a traditional sense. 

We also have shown that there are three \emph{incorrect} four-vector theorems: (i) M\o ller's theorem, which is also called ``M\o ller's version of Laue's theorem'' in Ref.\,\cite{r6}; (ii) Landau-Lifshitz version of Laue's theorem; and (iii) Weinberg's version of Laue's theorem. M\o ller's version is disproved in the present paper by taking the mathematical tensor Eq.\,(\ref{eq38}) as a counterexample, while Landau-Lifshitz and Weinberg's versions are disproved in Ref.\,\cite{r6} by taking the EM tensor of a charged metal sphere in free space as a counterexample. All the sufficient conditions of the three \emph{disproved} versions of Laue's theorem include the divergence-less of a tensor ($\partial_{\nu}\Theta^{\mu\nu}(\mathbf{x},t)=0$), but they do not include or derive the sufficient and necessary condition Eq.\,(\ref{eq10}). Thus it is not appropriate that $\partial_{\nu}\Theta^{\mu\nu}(\mathbf{x},t)=0$ is recognized to be ``conservation Law'' \cite[p.\,310]{r18} or to ``guarantee conservation of the total 4-momentum'' \cite[p.\,443]{r21} in traditional textbooks.

The most convincing way to disprove a mathematical conjecture is to provide its counterexample.  The counterexample Eq.\,(\ref{eq38}) for M\o ller's version of Laue's theorem, $A^{\mu\nu}$, is also a counterexample of Landau-Lifshitz and Weinberg's versions of Laue's theorem, because $A^{\mu\nu}$ is both divergence-less ($\partial_{\nu}A^{\mu\nu}=0$) and symmetric ($A^{\mu\nu}=A^{\nu\mu}$), while the Landau-Lifshitz version takes the divergence-less of a tensor as a sufficient condition, and the Weinberg's version takes the divergence-less plus a symmetry of a tensor as a sufficient condition. In other words, $A^{\mu\nu}$ given by Eq.\,(\ref{eq38}) is a common \emph{mathematical} counterexample to disprove M\o ller's, Landau-Lifshitz, and Weinberg's versions of Laue's theorem which are all \emph{mathematical} conjectures, independent of physics although originating from physics, because all physical implications have already been turned into mathematical descriptions. 

There is a well-known result in the dynamics of relativity that if a Lorentz four-vector is divergence-less, then the time-element space integral of the four-vector is a Lorentz scalar; for example, Weinberg claims that ``for any conserved four-vector'', namely for any four-vector  $J^{\mu}=(\mathbf{J},c\rho)$ that satisfies equation $\partial_{\mu}J^{\mu}=0$, $cQ=\int J^4\mathrm{d}^3x=\int c\rho\mathrm{d}^3x$ ``defines a time-independent scalar'' \cite[p.41]{r2}, and M\o ller also claims a proof of such a result in his textbook \cite[p.168]{r3}.  However in the present paper, we have shown based on Theorem 3 in Sec.\,\ref{s4} that this well-known result is not correct.  We have found a general conclusion for the Lorentz property of a four-vector's time-element space integral, stating that whether the time-element space integral of a four-vector is a Lorentz scalar has nothing to do with the four-vector's divergence-less property. Accordingly, it is not appropriate for the current continuity equation $\partial_{\mu}J^{\mu}(\mathbf{x},t)=0$  to be called ``invariant conservation Law'' \cite[p.\,40]{r2} or ``the law of charge conservation'' \cite[p.\,559,\,p.\,443]{r21} in textbooks.

As presented in traditional textbooks, the \emph{local} conservation law of energy--momentum in general relativity, given by $T^{\mu\nu}_{~~~;\nu}=0$ \cite[p.\,298]{r22}, is covariantly generalized from the conservation law $T^{\mu\nu}_{~~~,\nu}=\partial_{\nu}T^{\mu\nu}=0$  in special relativity \cite[p.\,83]{r1}\,\cite[p.\,45]{r2}\,\cite{r26}.  Thus the validity of the latter is a necessary condition for the validity of the former.  This law is often used for relativistic analysis of the Abraham-Minkowski debate on the momentum of light in a medium \cite{r9,r10,r11}, and it is also thought to play an important role in gravitation theory \cite[p.132]{r21}, because ``the GR [general relativity] field equations should be consistent with energy and momentum conservation, $T^{\mu\nu}_{~~~;\nu}=0$'' \cite[p.\,299]{r22}. This law is so well-established that often no citations are given for its origin in research articles \cite{r27,r28,r29,r30}.  However in fact, the traditional conservation laws including both $\partial_{\mu}\mathnormal{\Lambda}^{\mu}=0$ and $\partial_{\nu}T^{\mu\nu}=0$ \cite[p.\,443]{r21} can be directly disproved by simple physical counterexamples, as shown in Appendix \ref{appc}, although they are claimed to have been already proved with the use of the modern language of exterior calculus \cite[p.\,318]{r19}. Thus clarifying the two conservation laws in the present paper has a general significance.

In conclusion, we have generally resolved two fundamental issues in the dynamics of relativity: (a) Under what condition, the time-column space integrals of a Lorentz four-tensor constitute a Lorentz four-vector (Theorem 1), and (b) under what condition, the time-element space integral of a Lorentz four-vector is a Lorentz scalar (Theorem 3).  Both Theorem 1 and Theorem 3 have their own sufficient and necessary conditions, which have nothing to do with the divergence-less ($\partial_{\nu}\Theta^{\mu\nu}=0$ or $\partial_{\mu}\mathnormal{\Lambda}^{\mu}=0$), symmetry ($\Theta^{\mu\nu}=\Theta^{\nu\mu}$), and boundary conditions.  This point is especially important.  For example, from Theorem 1 we can directly judge that M\o ller's theorem is incorrect, because the sufficient condition of M\o ller's theorem does not include the sufficient and necessary condition of Theorem 1, as shown in counterexample Eq.\,(\ref{eq38}).  A similar argument is applicable to the ``invariant conservation law'' claimed by Weinberg \cite[p.\,40]{r2}, as shown in counterexample Eq.\,(\ref{eq48}). 

As a practical application, we have used Theorem 1 to analyze Minkowski EM tensor for a plane light wave in a moving medium with Einstein's light--quantum hypothesis taken into account, and we find that the four-momentum of quasi-photon and the Lorentz invariance of Planck constant can be naturally derived (see Appendix \ref{appb}).  Einstein's light--quantum hypothesis is the basis of Bohr frequency condition of radiation from atomic transitions, the invariance of Planck constant is an implicit postulate in Dirac relativistic quantum mechanics \cite{r7}, and the existence of four-momentum of the quasi-photon is required by momentum--energy conservation law in Einstein-box thought experiment \cite{r25}.  Thus this natural derivation is compatible with the quantum theory and the momentum--energy conservation law, and the result obtained strongly supports the Minkowski tensor as being the most qualified momentum--energy tensor for descriptions of light--matter interactions in the frame of the principle of relativity. 

In relativity, charge conservation law refers to that the total charge is a \emph{time-} and \emph{frame-independent} constant \cite[p.\,41]{r2}, as demonstrated in Eq.\,(\ref{eqa6}) of Appendix \ref{appa}, while energy-momentum conservation law refers to that the total energy and momentum constitute a \emph{covariant} and \emph{time-independent} four-vector \cite[p.\,46]{r2}, as demonstrated for a plane light wave in Appendix \ref{appb}.  The problems of $\partial_{\mu}\mathnormal{\Lambda}^{\mu}=0$  and $\partial_{\nu}\Theta^{\mu\nu}=0$  as being conservation laws in relativity were first discovered in \cite{r6,r7}, and they are explicitly illustrated and generally resolved in the present paper.  One might argue that in the establishment of the positive mass theorems in general relativity, rigorous mathematics has been used and loopholes possibly left in earlier works do not exist.  Unfortunately, this is not true.   These problems \emph{never} got any attention in the proofs of the positive mass theorem by Schoen and Yau in 1979 \cite{r32} and 1981 \cite{r39}, Witten \cite{r33} and Nester \cite{r34} in 1981, Parker and Taubes in 1982 \cite{r35}, and Gibbons and coworkers in 1983 \cite{r36}.  All the proofs are based on a flawed theoretical framework set up by Arnowitt, Deser, and Misner \cite{r37,r38}, where the total energy--momentum $P^{\mu}$ in an asymptotically flat spacetime is required to ``obey the differential conservation law  $T^{\mu\nu}_{~~~,\mu}=0$ [of the energy--momentum tensor $T^{\mu\nu}$]'' and required to ``transform as a four-vector'' \cite{r38}. However in fact, the conservation law  $T^{\mu\nu}_{~~~,\mu}=0$ cannot guarantee that $P^{\mu}$  is a four-vector, as shown in Appendix \ref{appd}. Apparently, this problem is ignored in all the proofs \cite{r32,r33,r34,r35,r36,r39}.  Moreover, the equations  $T_{\mu~;\nu}^{~\nu}=0$,  $T^{\mu\nu}_{~~~;\mu}=0$, $T^{\mu\nu}_{~~~;\nu}=0$, or $T^{\mu\nu}_{~~~,\nu}=0$  have been always claimed as conservation law in later literature, such as in Ref.\,\cite{r28} by Ratra and Peebles in 1988, Ref.\,\cite{r27} by Witten in 1991, Ref.\,\cite[p.\,299]{r22} by Rindler in 2006, and Ref.\,\cite{r29} by Bi\v{c}\'{a}k and Schmidt in 2016.  From this one can see that the community in the field of relativity has never become aware of these problems before the publications of \cite{r6,r7}. 

Recently, Bojowald criticizes that the constant wave four-vector $K^{\mu}$  in Appendix \ref{appc} does not constitute a counterexample to the traditional conservation law $\partial_{\mu}\mathnormal{\Lambda}^{\mu}=0\Rightarrow\int\mathnormal{\Lambda}^4\mathrm{d}^3x=\mathrm{invariant}$, because ``it does not obey the usual boundary conditions of conservation laws, which state that the integrated current must vanish at timelike boundaries or at infinity'' \cite{r40}.  Apparently, Bojowald's criticism has no basis at all, because the conservation law $\partial_{\mu}\mathnormal{\Lambda}^{\mu}=0$, as claimed in \cite[p.\,40]{r2}\cite{r19}\cite[p.\,443]{r21}, does not require $\mathnormal{\Lambda}^{\mu}=0$ holding on boundaries.  A typical example is the current density four-vector $J^{\mu}=(\mathbf{J},c\rho)$ for a sphere of perfect conductor with a radius of $R~(\ne 0)$  and a total static charge of $Q~(\ne 0)$ in free space.  Observed in the sphere-rest frame, the current density is given by $\mathbf{J}=0$ holding everywhere, and according to the basic properties for a perfect conductor, we know that any net charge resides on the surface \cite[p.\,97]{r41}, with the charge density given by  $\rho=Q\delta (r-R)/(4\pi R^2)$, where $\delta(x)$  is Dirac delta function.  The current four-vector $J^{\mu}$  satisfies the invariant conservation law $\partial_{\mu}J^{\mu}=0$  with $cQ=\int J^4\mathrm{d}^3x$  as a Lorentz invariant, as claimed by Weinberg \cite[p.\,40]{r2};  however, $J^4=c\rho=cQ\delta (r-R)/(4\pi R^2)\ne 0$ holds on the spherical \emph{boundary}  $r=R$.  From this elegant example of classical electrodynamics --- four-current $J^{\mu}$  not vanishing on boundary for charged metallic sphere in free space --- one can see that Bojowald's criticism is exactly based on his groundless speculation, and it is not justified. Thus $K^{\mu}\ne 0$ on boundaries does not lose its qualification to be a counterexample; namely, the wave four-vector $K^{\mu}$  with $\partial_{\mu}K^{\mu}=0$  and $\int K^4\mathrm{d}^3x\ne \mathrm{invariant}$  in Appendix \ref{appc} is a completely qualified counterexample to the traditional conservation law $\partial_{\mu}\mathnormal{\Lambda}^{\mu}=0\Rightarrow\int\mathnormal{\Lambda}^4\mathrm{d}^3x=\mathrm{invariant}$.

Bojowald also criticizes that instead of from Theorem 1 developed in the present paper, the total momentum--energy four-vector should be constructed from a covariant combination of a stress--energy tensor and the hyperplane differential-element (HDE) four-vector \cite{r40}, as presented in literature such as the textbook by Jackson \cite[p.\,757]{r4}.  However as shown in Appendix \ref{appa}, the HDE four-vector is apparently in contradiction with the principle of relativity and the principles of classical mathematical analysis.  Thus this criticism by Bojowald is not justified either, coming from his negligence of the self-consistency of a physical theory, and his lack of a good knowledge about how to transform a space integral from one inertial frame to another according to the \emph{change of variables theorem} in mathematical analysis \cite{r14}.


\appendix
\section{\label{appa} How to transform space integrals between Lorentz inertial frames?}
In special relativity, the hyperplane differential-element (HDE) four-vector is often used to transform triple (space) integrals from one Lorentz inertial frame to another \cite[p.\,757]{r4}.  In this Appendix, we will show that the HDE four-vector itself is not consistent with the principles of classical mathematical analysis and the principle of relativity although it is Lorentz covariant (confer Appendix \ref{appe} for the definition of covariance). 

Under time-space Lorentz transformation, there are two techniques used to transform space (triple) integrals between inertial frames.  The first technique is based on the \emph{change of variables theorem}, as presented in mathematical analysis \cite{r14}, and used in Laue's original paper to derive Laue's theorem \cite{r12}, and also used to develop my theory in the present paper and the previous work \cite{r6}; called Technique-I for convenience. In this technique, the differential element transformation formula or the \emph{change of variable formula} from the laboratory frame $XYZ$ to the frame $X'Y'Z'$ is given by  
\begin{align}
\mathrm{d}^3x&=\left|\frac{\partial(x,y,z)}{\partial(x',y',z')}\right|\mathrm{d}^3x'
=\gamma\mathrm{d}^3x', 
\label{eqa1}
\end{align} 
where $X'Y'Z'$ moves at $\mb{\beta}c$ with respect to $XYZ$, while $XYZ$  moves at $\mb{\beta}'c=-\mb{\beta}c$  with respect to  $X'Y'Z'$; $\mathrm{d}^3x$  is the proper differential element fixed in  $XYZ$, while $\mathrm{d}^3x'$ is the corresponding element moving at $\mb{\beta}'c=-\mb{\beta}c$ in  $X'Y'Z'$; Jacobian determinant $\partial(x,y,z)/\partial(x',y',z')=\gamma$  with $\gamma=(1-\mb{\beta}^2)^{-1/2}=(1-\mb{\beta}'^2)^{-1/2}=\gamma'$  is calculated from Eq.\,(\ref{eq3}), as illustrated in the \emph{Analytical example} in Sec.\,\ref{s2}. Note that $\gamma\mathrm{d}^3x'$  (=moving element $\mathrm{d}^3x'$ multiplied by its relativistic factor $\gamma'=\gamma$) is a Lorentz invariant in the Technique-I.

The second technique is based on the HDE (hyperplane differential-element) four-vector, as presented in some respected textbooks, such as the book by Jackson \cite[p.\,757]{r4}; called Technique-II for convenience.  In Technique-II, the integral domain $V$ is assumed to be at rest in the laboratory frame, and then similar to the classical particle's four-momentum = proper mass multiplied by its four-velocity, a differential-element four-vector (= proper element multiplied by its four-velocity normalized to light speed $c$) is constructed to define the change of variable formula. In the laboratory frame $XYZ$, the HDE four-vector is given by 
\begin{align}
\mathrm{d}S^{\mu}&=(\mathrm{d}\mathbf{S}, \mathrm{d}S^4)
\notag \\
&=(\mathbf{0},1)\mathrm{d}^3x=(0,0,0,1)\mathrm{d}^3x.
\label{eqa2}
\end{align}  
In the frame  $X'Y'Z'$ moving at $\mb{\beta}c$ with respect to $XYZ$, the HDE four-vector is given by
\begin{align}
\mathrm{d}S'^{\mu}=(\mathrm{d}\mathbf{S}', \mathrm{d}S'^4)=(\gamma'\mb{\beta}',\gamma')\mathrm{d}^3x, 
\label{eqa3}
\end{align} 
where $(\gamma'\mb{\beta}',\gamma')$  is the normalized four-velocity of $V$ (fixed in $XYZ$)  moving at $\mb{\beta}'c=-\mb{\beta}c$ observed in $X'Y'Z'$.  With $\gamma'=\gamma$ taken into account we have
\begin{equation}
 \mathrm{d}S'^4=\mathrm{d}^3x'=\gamma\mathrm{d}^3x=\gamma\mathrm{d}S^4 
\label{eqa4}
\end{equation} 
---\,the \emph{change of variable formula} for Technique-II.  In above Eqs.\,(\ref{eqa2}) and (\ref{eqa3}), $\mathrm{d}S^{\mu}=(\mathbf{0},1)\mathrm{d}^3x$  means that $\mathrm{d}^3x$  is fixed in  $XYZ$, while  $\mathrm{d}S'^{\mu}=(\mb{\beta}',1)\mathrm{d}^3x'$  means that  $\mathrm{d}^3x'= \gamma\mathrm{d}^3x$  moves at $\mb{\beta}'c=-\mb{\beta}c$ in  $X'Y'Z'$.  Note that $\mathrm{d}^3x'/\gamma$ $(=\mathrm{d}^3x)$ is  an invariant in the Technique-II.

In special relativity developed by Einstein \cite{r20}, a moving object Lorentz-contracts in the direction of motion, while keeping in the same the dimensions of the object in all the transverse directions.  Now $\mathrm{d}^3x$  is fixed in $XYZ$, and it is moving observed in $X'Y'Z'$.  Thus according to Einstein, $\mathrm{d}^3x'$, denoting the size of $\mathrm{d}^3x$ observed in $X'Y'Z'$, is given by
\begin{equation}
\mathrm{d}^3x'=\frac{\mathrm{d}^3x}{\gamma}, 
\label{eqa5}
\end{equation} 
which is exactly the same as  Eq.\,(\ref{eqa1}) derived from the change of variables theorem \cite{r14}. That is because the change of variables theorem is the strict mathematical basis of the effect of Lorentz contraction under time-space Lorentz transformation, as shown in the \emph{Analytical example} in Sec.\,\ref{s2}.

However according to Eq.\,(\ref{eqa4}), Technique-II requires $\mathrm{d}^3x'=\gamma\mathrm{d}^3x$; thus Technique-II contradicts both the change of variables theorem in mathematical analysis and and the effect of Lorentz contraction in Einstein's special relativity.  More seriously, Technique-II directly contradicts Lorentz invariance of total charge; in other words, if Technique-II were used, then a non-zero total charge would not be a Lorentz invariant, which is shown below.

\emph{Total charge Q = invariant in Technique-I.} First we show that Lorentz invariance of total charge is always valid in Technique-I.  Without loss of generality, we assume that a charge distribution is created by charged particles which move at the same velocity, otherwise it can be treated discretely, as shown in Ref.\,\cite{r6}.  According to special relativity, there must exist an inertial frame $XYZ$ where the charged particles are at rest.  Thus in the \emph{charge-rest} frame $XYZ$ (taken as the laboratory frame), the total charge can be formulated as  $Q=\int_V \rho(\mathbf{x})\mathrm{d}^3x$, where $\rho(\mathbf{x})$ with $\partial\rho/\partial t\equiv 0$ is the charge distribution, $V$  is the volume at rest, and $Q$  is the (time-independent) total charge in  $V$.  In such a case, all charged particles are stationary and frozen in $V$  so that no current exists ($\mathbf{J}=0$).  Observed in a frame $X'Y'Z'$ moving at $\mb{\beta}c\ne 0$  with respect to the charge-rest frame $XYZ$, the volume $V'$  is moving at   $\mb{\beta}'c=-\mb{\beta}c$, but there are no charged particles crossing through the boundary of $V'$  although the current $\mathbf{J}'=-\gamma\mb{\beta}c\rho\ne 0$  holds.  In Technique-I, as shown in Eq.\,(\ref{eqa5}), the change of variable formula is given by $\mathrm{d}^3x'=\mathrm{d}^3x/\gamma$.  From this, with $c\rho'=\gamma(c\rho-\mb{\beta}\cdot\mathbf{J})$ and $\mathbf{J}=0$ taken into account, we have 
\begin{align}
cQ':&=\int_{V'} (c\rho')\mathrm{d}^3x'
\notag \\
&=\int_V \gamma(c\rho-\mb{\beta}\cdot\mathbf{J})(\mathrm{d}^3x/\gamma)
\notag \\
&= \int_V (c\rho)\mathrm{d}^3x=:cQ
\notag \\
&=\mathrm{invariant}.
\label{eqa6}
\end{align} 
Thus we finish the proof that the total charge is always a Lorentz invariant in Technique-I. 

\emph{Total charge Q = non-invariant in Technique-II.} Now we show why Technique-II contradicts the Lorentz invariance of total charge.  In the laboratory frame $XYZ$ (charge-rest frame), the four-current density is given by $J^{\mu}=(\mathbf{J},J^4)$ with $\mathbf{J}=0$  and $J^4=c\rho$, and the total charge $Q$  is defined by  $cQ=\int_V(c\rho)\mathrm{d}^3x=\int_VJ^4\mathrm{d}S_4$.  Observed in the frame $X'Y'Z'$ moving at $\mb{\beta}c$  with respect to the laboratory frame $XYZ$, we have  $J'^{\mu}=(\mathbf{J}',J'^4)$  with $\mathbf{J}'=-\gamma\mb{\beta}J^4$  and  $J'^4=c\rho'=\gamma J^4$, and the total charge $Q'$  is defined by  $cQ'=\int_{V'}(c\rho')\mathrm{d}^3x'=\int_{V'}J'^4\mathrm{d}S'_4$.  According to Eq.\,(\ref{eqa2}) and Eq.\,(\ref{eqa3}) of Technique-II, we have $\mathrm{d}S_{\mu}=(-\mathrm{d}\mathbf{S},\mathrm{d}S_{4})$, with   $\mathrm{d}\mathbf{S}=0$ and $\mathrm{d}S_{4}=\mathrm{d}^3x$; and $\mathrm{d}S'_{\mu}=(-\mathrm{d}\mathbf{S'},\mathrm{d}S'_{4})=(\gamma\mb{\beta},\gamma)\mathrm{d}S_{4}$, with $\mathrm{d}\mathbf{S'}=-\gamma\mb{\beta}\mathrm{d}S_{4}$ and  $\mathrm{d}S'_{4}=\gamma\mathrm{d}S_{4}=\mathrm{d}^3x'$.  From this we have
\begin{align}
\int_V J^4\mathrm{d}S_4&=\int_V J^{\mu}\mathrm{d}S_{\mu}=\int_{V'} J'^{\mu}\mathrm{d}S'_{\mu} 
\notag \\
&=\int_{V'} J'^{4}\mathrm{d}S'_{4}-\int_{V'}\mathbf{J}'\cdot\mathrm{d}\mathbf{S}'  
\notag \\
&=\int_{V'} J'^{4}\mathrm{d}S'_{4}-\int_{V}(-\gamma\mb{\beta}J^4)\cdot(-\gamma\mb{\beta}\mathrm{d}S_4)  
\notag \\
&=\int_{V'} J'^{4}\mathrm{d}S'_{4}-(\gamma\mb{\beta})^2\int_{V}J^4\mathrm{d}S_4  \notag 
\end{align}
\begin{align}
&\Rightarrow~~\Big(1+(\gamma\mb{\beta})^2\Big)\int_{V}J^4\mathrm{d}S_4=\int_{V'} J'^{4}\mathrm{d}S'_{4} 
\notag \\ 
&\Rightarrow \quad \gamma^2cQ=cQ'  \quad\Rightarrow\quad  Q\ne Q' 
\label{eqa7}
\end{align}  \\
if $ Q\ne 0$ holds, where $\mb{\beta}\ne 0$ is assumed $\Rightarrow\gamma>1$. Thus we finish the proof that a non-zero total charge $(Q\ne 0)$ is not a Lorentz invariant in Technique-II.

So far we have shown that the total charge $Q\ne 0$  is never a Lorentz invariant in Technique-II, while the total charge $Q$  is always a Lorentz invariant in Technique-I; both cases have nothing to do with the boundary conditions of  $J^{\mu}=(\mathbf{J},c\rho)$.\vspace{0mm}

\emph{Conclusion.}  From above analysis we can conclude that Technique-II is based on the HDE four-vector, as presented in \cite[p.\,757]{r4}, and it has three flaws: (i) contradicting the \emph{change of variables theorem} in mathematical analysis \cite{r14}; (ii) contradicting the effect of Lorentz contraction in special relativity;  and (iii) contradicting the Lorentz invariance of total charge. To put it simply, Technique-II follows \emph{neither} the principles of mathematical analysis \emph{nor} the principle of relativity. Thus the Lorentz covariant  hyperplane differential-element four-vector Eq.\,(\ref{eqa3}) is only a \emph{mathematical} four-vector, instead of a \emph{physical} four-vector \cite{r24}.\vspace{0mm}

\emph{Two subtle, but apparently fundamental issues.}  In analysis of the Lorentz invariance of total charge in a volume, a subtle and important issue is about how to define the volume. If there are charged particles crossing through the boundary of the volume, the total charge in the volume may not be Lorentz invariant \cite{r15}, possibly leading to a doubt of the completeness of Maxwell EM theory \cite{r16}.  Thus the correct volume is supposed to be the one that moves at the same velocity as that of the charge, as argued above.  Another subtle issue is how to correctly understand the definition of total charge.  For example, by analysis of an infinite straight charged wire, Bili\`{c} puzzled that the standard definition $Q=\int_V\rho\mathrm{d}^3x$  and the so-called covariant definition $Q=\int J^{\mu}\mathrm{d}S_{\mu}$ (in units with $c=1$) are not equivalent in general \cite{r15}; now we know that the problem turned out to be here: the transform of triple integral  $\int J^{\mu}\mathrm{d}S_{\mu}=\int J'^{\mu}\mathrm{d}S'_{\mu}$ contradicts the \emph{change of variables theorem} in mathematical analysis \cite{r14}, as shown in Eqs.\,(\ref{eqa1}) and (\ref{eqa4}).  \vspace{0mm}

\section{\label{appb} Natural derivation of four-momentum of quasi-photon and invariance of Planck constant}
Quasi-photons are the carriers of the momentum and energy of the light in a medium, and they are descriptions of the macroscopic average of light-matter microscopic interactions \cite{r17}.  Lorentz invariance of Planck constant is an implicit postulate in Dirac relativistic quantum theory \cite{r7}. In this appendix, we will apply Theorem 1 to Minkowski EM tensor for a plane light wave in a moving uniform medium, and find that the momentum--energy four-vector (four-momentum) of the quasi-photon and the Lorentz invariance of Planck constant can be naturally derived.

For a plane light wave propagating in a moving non-dispersive, isotropic, uniform medium, observed in the \emph{medium-rest} frame, Minkowski quasi-photons characterizing the light-matter interactions move along the wave vector $\mathbf{k}_{\mathrm{w}}$ at the speed $c/n$, with $n ~(>1)$ the refractive index. In such a case, there is a \emph{photon-rest} frame moving at the quasi-photon velocity relatively to the medium-rest frame \cite{r7,r17}. Observed in the photon-rest frame $XYZ$ (taken as the laboratory frame here), there are some fantastic EM phenomena to take place. 
\begin{enumerate}
\item The EM fields $\mathbf{E}=0$, $\mathbf{H}=0$, $\mathbf{D}=\mathbf{D}_0\cos\Psi$, and $\mathbf{B}=\mathbf{B}_0\cos\Psi$  hold, where $\mathbf{D}_0\neq 0$  and  $\mathbf{B}_0\neq 0$ with $\mathbf{D}_0\,\bot\,\mathbf{B}_0$ are the constant amplitudes, leading to EM energy density  $W_{\mathrm{em}}=0.5(\mathbf{D}\cdot\mathbf{E}+\mathbf{B}\cdot\mathbf{H})=0$ and Abraham momentum  $\mathbf{g}_{\mathrm{A}}=\mathbf{E}\times\mathbf{H}/c^2=0$, but the Minkowski momentum  $\mathbf{g}_{\mathrm{M}}=\mathbf{D}\times\mathbf{B}\neq 0$. 

\item The wave angular frequency  $\omega=0$ and the wave phase $\Psi=(\omega t-\mathbf{k}_{\mathrm{w}}\cdot\mathbf{x})=(-\mathbf{k}_{\mathrm{w}}\cdot\mathbf{x})$  hold, with the wave vector $\mathbf{k}_{\mathrm{w}} \ne 0$, leading to all the EM fields behaving as \emph{static} fields \cite{r17}. 
\end{enumerate} 

The Minkowski EM stress--energy tensor $\widetilde{T}^{\mu\nu}$  is given by Eq.\,(\ref{eq43}), with $\check{\mathbf{T}}_{\mathrm{M}}=-\mathbf{DE}-\mathbf{BH}+\check{\mathbf{I}}W_{\mathrm{em}}=0$  and $c\mathbf{g}_{\mathrm{A}}=0$  holding $\Rightarrow$ the holding of $\int_{V:~t=const}\check{\mathbf{T}}_{\mathrm{M}}\mathrm{d}^3x=0$  and $\int_{V:~t=const}c\mathbf{g}_{\mathrm{A}}\mathrm{d}^3x=0$ $\Rightarrow$ the holding of $\int_{V:~t=const}\widetilde{T}^{\mu j}\mathrm{d}^3x=0$  for $\mu=1,2,3,4$  and $j=1,2,3$.  According to the sufficient and necessary condition Eq.\,(\ref{eq10}) of  Theorem 1,  $\widetilde{T}^{\mu\nu}$-time-column (= $T^{\mu\nu}$-time-row) space integrals 
\begin{align}
P^{\mu}=c\int_{V:~t=const}\left(~\mathbf{g}_{\mathrm{M}}(\Psi), ~\frac{W_{\mathrm{em}}(\Psi)}{c}~\right)\mathrm{d}^3x
\label{eqb1}
\end{align} 
constitute a Lorentz four-vector, which is time-independent because of all the EM fields behaving as being static in the photon-rest frame $\Rightarrow\partial\widetilde{T}^{\mu\nu}/\partial t\equiv 0$, including   $\partial \big(\mathbf{g}_{\mathrm{M}}, W_{\mathrm{em}}/c\big)/\partial t\equiv 0$. 

Observed in the inertial frame $X'Y'Z'$  moving relatively to the photon-rest frame $XYZ$  at a velocity of $\mb{\beta}c$, we have  
\begin{equation}
P'^{\mu}=c\int_{V':~t'=const}\left(~\mathbf{g}'_{\mathrm{M}}(\Psi'), ~\frac{W'_{\mathrm{em}}(\Psi')}{c}~\right)\mathrm{d}^3x',
\label{eqb2}
\end{equation}  
where the phase is given by  $\Psi'=(\omega' t'-\mathbf{k}'_{\mathrm{w}}\cdot\mathbf{x}')$, and $V'$  is moving at  $\mb{\beta}'c=-\mb{\beta}c$.  

From the Lorentz transformation of EM field-strength tensors \cite[see Eqs.\,(3) and (4) there]{r7}, with $\mathbf{E}=0$  and $\mathbf{H}=0$  taken into account we have 
\begin{align}
\mathbf{E}'&=\gamma\mb{\beta}\times\mathbf{B}c, 
\label{eqb3} 
\\
\mathbf{H}'&=-\gamma\mb{\beta}\times\mathbf{D}c, 
\label{eqb4} 
\\
\mathbf{D}'&=\gamma\mathbf{D}-\xi(\mb{\beta}\cdot\mathbf{D})\mb{\beta},
\label{eqb5} 
\\
\mathbf{B}'&=\gamma\mathbf{B}-\xi(\mb{\beta}\cdot\mathbf{B})\mb{\beta},
\label{eqb6} 
\end{align}
where $\gamma=(1-\mb{\beta}^2)^{-1/2}$  and $\xi=(\gamma-1)/\mb{\beta}^2$.  From above Eqs.\,(\ref{eqb3})-(\ref{eqb6}), we obtain 
\begin{equation}
\frac{\mathbf{E}'\times\mathbf{H}'}{W'_{\mathrm{em}}}=-\mb{\beta}c=\mb{\beta}'c,
\label{eqb7}
\end{equation} 
called the energy velocity traditionally, where $W'_{\mathrm{em}}=0.5(\mathbf{D'}\cdot\mathbf{E'}+\mathbf{B'}\cdot\mathbf{H'})=\mathbf{D'}\cdot\mathbf{E'} =\mathbf{B'}\cdot\mathbf{H'}$  is the EM energy density.  Because $(\mathbf{E}'\times\mathbf{H}')/W'_{\mathrm{em}}=\mb{\beta}'c$  given by above Eq.\,(\ref{eqb7}) is the moving velocity of the volume  $V'$, seemingly it indeed looks like the velocity of energy transport.  However in fact, it is not the real energy velocity in general.  The real energy velocity is the phase velocity, required by both Fermat's principle and energy conservation law which are physical postulates independent of Maxwell equations \cite{r23}.  That is why $\mb{\beta}'c=(\mathbf{E}'\times\mathbf{H}')/W'_{\mathrm{em}}$   is called ``photon apparent velocity'' in Ref.\,\cite{r7}.

According to the Lorentz transformation of the wave four-vector  $K^{\mu}=(\mathbf{k}_{\mathrm{w}},\omega/c)$, with $\omega=0$  in $XYZ$  taken into account, we have 
\begin{equation}
\omega=\gamma(\omega'-\mathbf{k}'_{\mathrm{w}}\cdot\mb{\beta}'c)~~\Rightarrow~~\omega'=\mathbf{k}'_{\mathrm{w}}\cdot\mb{\beta}'c.
\label{eqb8} 
\end{equation} 

As mentioned above, both Fermat's principle and energy conservation law require the EM energy to propagate at the phase velocity \cite{r7}, given by 
\begin{equation}
\mb{\beta}'_{\mathrm{ph}}c=\frac{\omega'}{|\mathbf{k}'_{\mathrm{w}}|}\frac{\mathbf{k}'_{\mathrm{w}}}{|\mathbf{k}'_{\mathrm{w}}|}~~\Rightarrow~~\omega'=\mathbf{k}'_{\mathrm{w}}\cdot\mb{\beta}'_{\mathrm{ph}}c.
\label{eqb9}
\end{equation} 
From Eq.\,(\ref{eqb8}) and Eq.\,(\ref{eqb9}), we have 
\begin{equation}
\mathbf{k}'_{\mathrm{w}}\cdot\mb{\beta}'=\mathbf{k}'_{\mathrm{w}}\cdot\mb{\beta}'_{\mathrm{ph}}.
\label{eqb10}
\end{equation}	\\
\indent As indicated, the volume $V'$  in  Eq.\,(\ref{eqb2}) moves at $\mb{\beta}'c$.  If the same volume $V'$  moves at  $\mb{\beta}'_{\mathrm{ph}}c$, we label it as  $V'_{\mathrm{light}}$, called ``light volume'' \cite{r7}.  Observed in the photon-rest frame  $XYZ$, the volume $V$ is at rest, and the light volume $V_{\mathrm{light}}$  is also at rest due to $\omega=0 \Rightarrow \mb{\beta}_{\mathrm{ph}}c=0$; namely $V_{\mathrm{light}}=V$.

Observed in  $X'Y'Z'$, from Eq.\,(\ref{eqb10}) we find that $\Psi'=(\omega' t'-\mathbf{k}'_{\mathrm{w}}\cdot\mathbf{x}')$  takes the same value for $\mathbf{x}'=\mathbf{x}'_0+\mb{\beta}'ct'$  with $\mathbf{x}'\in V'$  and for $\mathbf{x}'=\mathbf{x}'_0+\mb{\beta}'_{\mathrm{ph}}ct'$  with $\mathbf{x}'\in V'_{\mathrm{light}}$.  Thus with  $V'$  in Eq.\,(\ref{eqb2}) replaced by $V'_{\mathrm{light}}$, we obtain an equivalent expression, given by \\
\begin{equation}
P'^{\mu}=c\int_{V'_{\mathrm{light}}:~t'=const}\left(~\mathbf{g}'_{\mathrm{M}}(\Psi'), ~\frac{W'_{\mathrm{em}}(\Psi')}{c}~\right)\mathrm{d}^3x'.
\label{eqb11}
\end{equation} \\
The above Eq.\,(\ref{eqb11}) is a time-independent and Lorentz covariant four-vector. 

So far, with Theorem 1 applied to Minkowski EM tensor we have arrived at a conclusion that for a plane light wave in a medium, the total Minkowski EM momentum and energy contained in a \emph{light volume}  constitute a Lorentz four-vector. 

The above result implies that observed in any inertial frame, all photons in a light volume are moving at the same velocity as that of the light volume and no photons cross the boundary of the volume; in other words, observed at the same time in all inertial frames, respectively, the photons in the light volume are the same photons.  This Lorentz property of light volume was first formulated in Ref.\,\cite[see Eq.\,(50) there]{r7}, directly based on the Lorentz covariance of EM field-strength tensors (instead of Minkowski EM tensor). This makes sense, because Minkowski tensor is a covariant combination of the EM field-strength tensors \cite[footnote 7]{r17}, and in principle, all physical results obtained from Minkowski tensor are already embodied in the EM field-strength tensors \cite{r24}.

Especially, if there is only one photon contained in the light volume $V'_{\mathrm{light}}$ and Einstein light-quantum hypothesis is taken into account, namely 
\begin{equation}
\int_{V'_{\mathrm{light}}: ~t'=const} W'_{\mathrm{em}}\mathrm{d}^3x'=\hbar\omega', 
\label{eqb12}
\end{equation} 
then  
\begin{align}
\int_{V'_{\mathrm{light}}: ~t'=const} \mathbf{g}'_{\mathrm{M}}\mathrm{d}^3x'&=\int_{V'_{\mathrm{light}}: ~t'=const} (\mathbf{D}'\times\mathbf{B}')\mathrm{d}^3x'
\notag \\
&=\int_{V'_{\mathrm{light}}: ~t'=const} (W'_{\mathrm{em}}/\omega')\mathbf{k}'_{\mathrm{w}}\mathrm{d}^3x'
\notag \\
\notag \\
&=\hbar\mathbf{k}'_{\mathrm{w}}
\label{eqb13}
\end{align}
is the momentum of the photon in  $V'_{\mathrm{light}}$,  where the Lorentz invariant expression $\mathbf{D}'\times\mathbf{B}'=(W'_{\mathrm{em}}/\omega')\mathbf{k}'_{\mathrm{w}}$  is used \cite[see Eq.\,(37) there]{r7}, and $\hbar$  is the reduced Planck constant.  Inserting Eq.\,(\ref{eqb12}) and Eq.\,(\ref{eqb13}) into Eq.\,(\ref{eqb11}), we find that
\begin{equation}
(\hbar\mathbf{k}'_{\mathrm{w}},\hbar\omega'/c)=\hbar K'^{\mu}\quad(=P'^{\mu}/c)
\label{eqb14}
\end{equation} 
is the photon's four-momentum, where $K'^{\mu}=(\mathbf{k}'_{\mathrm{w}},\omega'/c)$  is a known (wave) four-vector.  Since $\hbar K'^{\mu}$  and  $K'^{\mu}$  are both four-vectors, $\hbar(g_{\mu\nu}K'^{\mu}X'^{\nu})=scalar$  and $(g_{\mu\nu}K'^{\mu}X'^{\nu})=scalar$  must hold, where  $g_{\mu\nu}$ is the Minkowski metric and $X'^{\nu}$  is the time-space four-vector.  Note that $(g_{\mu\nu}K'^{\mu}X'^{\nu})$  is the scalar of phase, and it can be any real number.  Thus the Planck constant $\hbar$  must be a Lorentz invariant. This conclusion was first obtained in Ref.\,\cite{r7} directly from two EM field-strength tensors.  

\emph{Conclusion.} From above it is seen that the four-momentum of quasi-photon in a medium and the invariance of Planck constant are naturally obtained by applying Theorem 1 to Minkowski tensor, with taken into account Einstein light-quantum hypothesis that is the basis of Bohr frequency condition of atomic transitions in quantum theory.  On the other hand, as shown in Einstein-box thought experiment \cite{r25}, momentum--energy conservation law requires the quasi-photon to have a four-momentum. Thus the Minkowski tensor is compatible with the quantum theory of atomic light radiation and the momentum--energy conservation law, and it is the correct momentum--energy tensor for descriptions of light--matter interactions. \vspace{1.5mm}

\emph{Question 1:}  Is the generalized von Laue's theorem \cite{r6} applicable for identifying the Lorentz property of light momentum and energy for a plane light wave in a medium?  The answer is yes, because its pre-assumption $\partial\widetilde{T}^{\mu\nu}/\partial t\equiv 0$  is satisfied, as shown above. \vspace{1.5mm}
 
\emph{Question 2:}  Is the corrected M\o ller's theorem applicable for identifying the Lorentz property of light momentum and energy for a plane light wave in a medium?  The answer is no, because this plane light wave is a non-trivial solution of Maxwell equations (non-zero field solution), and the pre-assumption of corrected M\o ller's theorem cannot be satisfied, namely the divergence-less $\partial_{\nu}\widetilde{T}^{\mu\nu}=\partial_{\nu}T^{\nu\mu}=\big(\nabla\cdot\check{\mathbf{T}}_{\mathrm{M}}+\partial\mathbf{g}_{\mathrm{M}}/\partial t,~\nabla\cdot(c\mathbf{g}_{\mathrm{A}})+\partial W_{\mathrm{em}}/\partial (ct)\big)=0$ is fulfilled \cite{r6}, but the zero-boundary condition ($\widetilde{T}^{\mu\nu}=0$ on the boundary of $V$) cannot be fulfilled; for example, $(\widetilde{T}^{14},\widetilde{T}^{24},\widetilde{T}^{34})$ = $c\mathbf{g}_{\mathrm{M}}=c\mathbf{D}\times\mathbf{B}=c\mathbf{D}_0\times\mathbf{B}_0\cos^2(-\mathbf{k}_{\mathrm{w}}\cdot\mathbf{x})=0$ cannot hold on the \emph{whole closed boundary surface} of a finite 3D domain $V\neq 0$  because $\cos^2(-\mathbf{k}_{\mathrm{w}}\cdot\mathbf{x})=0$ only appears on the \emph{discrete planes} with $\mathbf{k}_{\mathrm{w}}\cdot\mathbf{x}=(2l+1)\pi/2$, where $l$  is an arbitrary integer.  \vspace{1.5mm}

\emph{Question 3:}  Is Theorem 1 applicable for identifying the Lorentz property of light momentum and energy for a plane light wave in \emph{free space}?  The answer is no.  

In free space, for a non-trivial plane wave observed in any given inertial frame $XYZ$, the EM fields can be written as  $(\mathbf{E},\mathbf{H},\mathbf{D},\mathbf{B})=(\mathbf{E}_0,\mathbf{H}_0,\epsilon_0\mathbf{E}_0,\mu_0\mathbf{H}_0)\cos\Psi$, where  $\mathbf{E}_0\neq 0$ and $\mathbf{H}_0\neq 0$  with  $\mathbf{E}_0\,\bot\,\mathbf{H}_0$  are the constant amplitudes, $\Psi=(\omega t-\mathbf{k}_{\mathrm{w}}\cdot\mathbf{x})$  is the wave phase, $\mathbf{k}_{\mathrm{w}}$  with  $|\mathbf{k}_{\mathrm{w}}|=\omega/c\ne 0$  is the wave vector, and $\epsilon_0$ and  $\mu_0$ the vacuum permittivity and permeability constants.  Abraham and Minkowski momentums are equal, namely $\mathbf{g}_{\mathrm{A}}=\mathbf{g}_{\mathrm{M}}=\mathbf{E}\times\mathbf{H}/c^2 ~(\ne 0)$.  

According to Theorem 1, the sufficient and necessary condition for the plane wave can be written as
\begin{align}
&\int_{V:~t=const}\widetilde{T}^{\mu j}(\mathbf{x},t)\mathrm{d}^3x=0 
\notag \\
&\hspace{18mm}\mathrm{for} \quad\mu=1,2,3,4 \quad\mathrm{and}\quad j=1,2,3,
\label{eqb15}
\end{align}				
where the volume $V$ is at rest in $XYZ$, and $\widetilde{T}^{\mu\nu}$ is given by Eq.\,(\ref{eq43}).  

From Eq.\,(\ref{eq43}) we have 
\begin{align}
&\int_{V:~t=const}(\widetilde{T}^{41},\widetilde{T}^{42},\widetilde{T}^{43})~\mathrm{d}^3x
\notag \\
&\hspace{25mm}=\int_{V:~t=const}c\mathbf{g}_{\mathrm{A}}\mathrm{d}^3x\ne 0  
\label{eqb16}
\end{align} \\
for a finite volume $V~(\neq 0)$.

From above Eq.\,(\ref{eqb16}) we find that the sufficient and necessary condition Eq.\,(\ref{eqb15}) cannot hold.  Thus we conclude that for a plane light wave in free space, the total (Minkowsi = Abraham) momentum and energy contained in a given volume $V$  that is at rest in an inertial frame cannot constitute a Lorentz four-vector. However, it does not mean that the momentum and energy of light cannot constitute a four-vector, because the photons in free space cannot be stopped in a given volume $V$  that is resting in an inertial frame in terms of Einstein's hypothesis of constancy of light speed.  From this it follows that always there are photons crossing through the boundary of  $V$, and flowing into and out of  $V$, and thus the photons in $V$  are not the same photons observed for different time. (Note that the photon density $N_{\mathrm{p}}=W_{\mathrm{em}}/(\hbar\omega)=[\epsilon_0E_0^2/(\hbar\omega)]\cos^2\Psi$  is a wave, dependent on time and space locations.)  On the other hand, because of the relativity of simultaneity, photons may cross through the boundary of $V$ at the same time in one frame, but these photons cannot cross through the boundary of  $V$  at the same time in other frames; thus leading to a result that the photons in  $V$  are not the same photons observed in different frames.  That is why the total momentum and energy of the photons contained in $V$  cannot constitute a four-vector in such a case. 

Therefore, Theorem 1 only can be used to identify the Lorentz property of the total momentum and energy of materials or particles, which are moving at a velocity less than the vacuum light speed $c$ so that there is a material-rest or particle-rest inertial frame, such as in the case for a plane light wave in a dielectric medium shown above, where Minkowski quasi-photon propagates at a velocity of  $c/n<c$ \cite{r17}. (Note: If a momentum--energy tensor is contributed by materials or particles which move at different velocities individually, then the tensor should be discretized so that each of the discretized tensors is contributed by the materials or particles which move at the same velocity, just like in the proof of the Lorentz invariance of total charge given in Ref.\,\cite{r6}).  

It should be noted that in free space, the invariance of Planck constant $\hbar$  and the covariance of wave four-vector  $K^{\mu}$ have been already proved in \cite{r7} and \cite{r20}, respectively.  Thus the four-momentum of photon in free space, equal to $\hbar K^{\mu}$, is a solved problem theoretically.\vspace{1.5mm}

\emph{Further specific explanation:} Why is 
\begin{align}
\int_{V:~t = const}c\mathbf{g}_{\mathrm{A}}\mathrm{d}^3x&=\int_{V:~t=const}\frac{\mathbf{E}\times\mathbf{H}}{c}~\mathrm{d}^3x
\notag \\
\notag \\
&\ne 0 
\notag 
\end{align} 
always valid for a finite $V\neq 0$  for a plane wave in free space?  

For a non-trivial plane wave in free space, observed in any inertial frames, the power flow or Poynting vector   
\begin{align}
&\mathbf{E}\times\mathbf{H}=\mathbf{E}_0\times\mathbf{H}_0\cos^2\Psi\not\equiv 0 
\notag \\
\Rightarrow\hspace{10mm} &\mathbf{E}_0\times\mathbf{H}_0\neq 0  
\notag  
\end{align} 
holds; otherwise, there are no energy flowing and no wave.  

On the other hand, we have $\int_{V:~t=const}\cos^2\Psi\mathrm{d}^3x>0$ holding; thus leading to the holding of $\int_{V:~t=const}c\mathbf{g}_{\mathrm{A}}\mathrm{d}^3x\neq 0$ for a plane wave in free space. 

Note: $\int_{V:~t=const}\cos^2\Psi\mathrm{d}^3x>0$ comes from the fact that $V\neq 0$ is a finite 3D volume, and $\cos^2\Psi\ge 0$  holds with the zero points only appearing on discrete planes, and thus there must exist a smaller volume $V^*\subset V$, where $\cos^2\Psi>0$ exactly holds $\Rightarrow$ the holding of $\int_{V:~t=const}\cos^2\Psi\mathrm{d}^3x\ge\int_{V^*:~t=const}\cos^2\Psi\mathrm{d}^3x>0$. \\

\section{\label{appc} Physical counterexamples \\ of Thirring's claims}
The most effective and convincing way to disprove a claim is to give its counterexample. In this appendix, we will provide two \emph{physical} counterexamples of the claims made by Thirring.

In his book \cite{r19}, with the help of \emph{exterior calculus} Thirring claims:\vspace{1.5mm}

(i) $\partial_{\mu}\mathnormal{\Lambda}^{\mu}=0$  makes $\int\mathnormal{\Lambda}^4\mathrm{d}^3x$ be a Lorentz scalar (namely ``invariant conservation Law'' claimed by Weinberg \cite[p.\,40]{r2}); \vspace{1.0mm}

(ii) $\partial_{\nu}T^{\mu\nu}=0$ makes $\int T^{\mu 4}\mathrm{d}^3x$ be a Lorentz four-vector (namely, Landau-Lifshitz version of Laue's theorem \cite{r6}). \vspace{2.5mm}

In both claims (i) and (ii), no boundary conditions are required.  However, claim (i) can be directly disproved by a simple counterexample $\mathnormal{\Lambda}^{\mu}=K^{\mu}$, and claim (ii) can be directly disproved by a simple counterexample $T^{\mu\nu}=K^{\mu}K^{\nu}$, where  $K^{\mu}=(\mathbf{k}_{\mathrm{w}},\omega/c)$ is the wave four-vector for a plane wave in \emph{free space}, first shown by Einstein \cite{r20}, with $\mathbf{k}_{\mathrm{w}}$  the wave vector, $\omega~ (\ne 0)$  the angular frequency, and $|\mathbf{k}_{\mathrm{w}}|=\omega/c$ holding.  This is illustrated as follows.  

Apparently, $\partial_{\mu}K^{\mu}=0$ and $\partial_{\nu}(K^{\mu}K^{\nu})=0$ are both valid because $K^{\mu}$  is independent of space and time variables $(\mathbf{x},t)$.  Suppose that $X'Y'Z'$ frame moves with respect to the laboratory frame $XYZ$  at a velocity of $\mb{\beta}c$ along the wave vector $\mathbf{k}_{\mathrm{w}}$.  Observed in  $XYZ$, we have $K^4=\omega/c$  and 
\begin{equation}
\int_V K^4\mathrm{d}^3x=\frac{\omega}{c}\int_V \mathrm{d}^3x,
\label{eqc1}
\end{equation}					
where the integral domain $V$  is fixed in  $XYZ$.    Observed in  $X'Y'Z'$, we have $K'^4=\omega'/c$  and 
\begin{equation}
\int_{V'} K'^4\mathrm{d}^3x'=\int_{V'}\frac{\omega'}{c}\mathrm{d}^3x',
\label{eqc2}
\end{equation}	
where the integral domain $V'$ is moving at $\mb{\beta}'c=-\mb{\beta}c$.

From the Lorentz transformation of $K^{\mu}$, we obtain the Doppler frequency shift \cite{r20}, given by 
\begin{align}
\frac{\omega'}{c}&=\gamma\left(\frac{\omega}{c}-\mb{\beta}\cdot\mathbf{k}_{\mathrm{w}}\right)
\notag \\
\notag \\
&=\gamma(1-|\mb{\beta}|)\frac{\omega}{c},
\label{eqc3}
\end{align} 
where $\gamma=(1-\mb{\beta}^2)^{-1/2}$  is the time dilation factor, and $\mb{\beta}\,\|\,\mathbf{k}_{\mathrm{w}}\Rightarrow\mb{\beta}\cdot\mathbf{k}_{\mathrm{w}}=|\mb{\beta}|\times |\mathbf{k}_{\mathrm{w}}|=|\mb{\beta}|\omega/c$ is employed.

As shown in Eq.\,(\ref{eqa5}) of Appendix \ref{appa}, the change of variable formula in such a case is given by $\mathrm{d}^3x'=\mathrm{d}^3x/\gamma$.  Inserting Eq.\,(\ref{eqc3}) into Eq.\,(\ref{eqc2}), with $\mathrm{d}^3x'=\mathrm{d}^3x/\gamma$  and Eq.\,(\ref{eqc1}) taken into account, we have 
\begin{align}
\int_{V'} K'^4\mathrm{d}^3x'&=\int_{V'} \frac{\omega'}{c}\mathrm{d}^3x'
\notag \\
&=\gamma(1-|\mb{\beta}|)\frac{\omega}{c}\int_V\frac{\mathrm{d}^3x}{\gamma}
\notag \\
&=(1-|\mb{\beta}|)\frac{\omega}{c}\int_V\mathrm{d}^3x
\notag \\
&=(1-|\mb{\beta}|)\int_V K^4\mathrm{d}^3x.
\label{eqc4}
\end{align}
Note that in above Eq.\,(\ref{eqc4}), the integrand $K'^{4}$  satisfies the definition  Eq.\,(\ref{eq24}) in the proof of Theorem 3. 

From above Eq.\,(\ref{eqc4}) we have 
\begin{equation}
\int_{V'} K'^4\mathrm{d}^3x'\ne \int_V K^4\mathrm{d}^3x ~~\mathrm{for} ~~\mb{\beta}c\ne 0.
\label{eqc5}
\end{equation} 
Thus we conclude that  $\int_V K^4\mathrm{d}^3x$ is not a Lorentz scalar, and Thirring's claim (i) is disproved. 

On the other hand, we have 
\begin{equation}
\int_V T^{\mu 4}\mathrm{d}^3x=\int_V K^{\mu}K^4\mathrm{d}^3x=K^{\mu}\int_V K^4\mathrm{d}^3x.
\label{eqc6}
\end{equation}  
In above Eq.\,(\ref{eqc6}), $K^{\mu}$ is a four-vector, but $\int_V K^4\mathrm{d}^3x$  is not a Lorentz scalar; thus their product 
\begin{equation}
(K^{\mu})\times\left(\int_V K^4\mathrm{d}^3x\right)=\int_V T^{\mu 4}\mathrm{d}^3x
\label{eqc7}
\end{equation} 
must not be a four-vector, and Thirring's claim (ii) is disproved as well.  \vspace{1.5mm}

\emph{Proof by contradiction that Eq.\,(\ref{eqc7}) is not a four-vector.}  If Eq.\,(\ref{eqc7}) were a four-vector, then 
\begin{equation}
(g_{\mu\nu}K^{\mu}X^{\nu})\left(\int_V K^4\mathrm{d}^3x\right) = \mathrm{scalar}
\label{eqc8}
\end{equation} 
would hold, where $X^{\nu}=(\mathbf{x},ct)$  is the time-space four-vector, and 
\begin{equation}
(g_{\mu\nu}K^{\mu}X^{\nu})=(\omega t-\mathbf{k}_{\mathrm{w}}\cdot\mathbf{x}) = \mathrm{scalar}
\label{eqc9}
\end{equation} 
is the scalar of phase, with $g_{\mu\nu}=\mathrm{diag}(-1,-1,-1,+1)$ the Minkowski metric. Thus from Eq.\,(\ref{eqc8}) and Eq.\,(\ref{eqc9}) it follows that $\big(\int_V K^4\mathrm{d}^3x\big)$  must be a scalar, but  $\big(\int_V K^4\mathrm{d}^3x\big)$  is not a scalar according to Eq.\,(\ref{eqc5}).  This contradiction indicates that  Eq.\,(\ref{eqc7}), $\int_V T^{\mu 4}\mathrm{d}^3x$=$K^{\mu}\big(\int_V K^4\mathrm{d}^3x\big)$, cannot be a Lorentz four-vector. \vspace{1.5mm}

\emph{An interesting question}: Why is $\int K^4\mathrm{d}^3x$  not a scalar for the wave four-vector  $K^{\mu}=(\mathbf{k}_{\mathrm{w}},\omega/c)$  while  $\int J^4\mathrm{d}^3x$  is a scalar for the current density four-vector  $J^{\mu}=(\mathbf{J},c\rho)$?  That is because the moving velocity of any charged particle is less than the vacuum light speed $c$, and there is a particle-rest frame where the particle current $\mathbf{J}=0\Rightarrow\int\mathbf{J}\mathrm{d}^3x=0$ holds, with the sufficient and necessary condition Eq.\,(\ref{eq23}) of Theorem 3 satisfied, $\Rightarrow\int J^4\mathrm{d}^3x=$ scalar, as shown in Eq.\,(\ref{eqa6}) of Appendix\,\ref{appa}.  However for  $K^{\mu}=(\mathbf{k}_{\mathrm{w}},\omega/c)$, there is no such a frame where $\mathbf{k}_{\mathrm{w}}=0\Rightarrow\int\mathbf{k}_{\mathrm{w}}\mathrm{d}^3x=0$ holds; thus $\int K^4\mathrm{d}^3x$  is not a scalar. Why is there no such a frame for  $\mathbf{k}_{\mathrm{w}} = 0$?  As we know, the photon momentum--energy four-vector is given by  $\hbar K^{\mu}=(\hbar\mathbf{k}_{\mathrm{w}},\hbar\omega/c)$, with $\hbar$  the Planck constant.  According to Einstein's hypothesis of constancy of light speed, there is no \emph{photon-rest} frame in free space, and the photon momentum $\hbar \mathbf{k}_{\mathrm{w}}\ne 0\Rightarrow\mathbf{k}_{\mathrm{w}}\ne 0$ holds in \emph{any} frames. 

\section{\label{appd} Why is the theoretical framework for the positive mass theorem flawed?}
In general relativity, the metric $g_{\mu\nu}$  are solutions of Einstein's field equations for a given energy--momentum tensor  $T^{\mu\nu}$ that causes space to curve \cite[p.\,5]{r21}. According to Arnowitt, Deser and Misner \cite{r37}, the definition of the total energy--momentum $P^{\mu}$  is given by the volume integral of the components of  $T^{0\mu}$, namely  
\begin{equation}
P^{\mu}:=\int T^{0\mu}\mathrm{d}^3x,  
\label{eqd1}
\end{equation} 
which can be expressed as surface integrals through Einstein's field equations and Gauss's theorem, and where $P^{0}=E$  is the total energy \cite{r38}.  

In the proofs of the positive mass theorem \cite{r33,r35,r39}, the total energy follows the definition in \cite{r37,r38}\,\cite[p.\,462]{r21}, given by 
\begin{equation}
E=\frac{1}{16\pi}\int (g_{jk,k}-g_{kk,j})\mathrm{d}^2S^j,
\label{eqd2}
\end{equation} 
where the integral is evaluated over a closed surface in the asymptotically flat region surrounding the source of gravitation.

Arnowitt, Deser and Misner claim that because of the conservation law $T^{\mu\nu}_{~~~,\mu}=0$, ``$P^{\mu}$  should transform as a four-vector'' \cite{r38}, which is clearly endorsed by Nester \cite{r34}. In the textbook by Misner, Thorne and Wheeler \cite[p.\,462]{r21}, it is also emphasized that the conservation law  $T^{\mu\nu}_{~~~,\mu}=0$ ($T^{\mu\nu}_{~~~,\nu}=0$) makes $P^{\mu}=\int T^{0\mu}\mathrm{d}^3x$ ($P^{\mu}=\int T^{\mu 0}\mathrm{d}^3x$) be a four-vector.  

Unfortunately, as shown by the counterexample in Appendix \ref{appc},  $T^{\mu\nu}_{~~~,\mu}=0$ cannot guarantee that $P^{\mu}$  is a four-vector.  Thus the Arnowitt-Deser-Misner theoretical framework used for the proofs of the positive mass theorem is flawed.  

Nevertheless, one might argue that in the framework of the positive mass theorem, in addition to $T^{\mu\nu}_{~~~,\mu}=0$  there is another important requirement, called \emph{dominant energy condition} \cite{r39}, reading: 
\begin{equation}
T^{00}\ge |T^{\mu\nu}|
\label{eqd3}
\end{equation} 
for each $\mu$, $\nu$ \cite{r30,r35}; and both $T^{\mu\nu}_{~~~,\mu}=0$ and  $T^{00}\ge|T^{\mu\nu}|$ together make $P^{\mu}=\int T^{0\mu}\mathrm{d}^3x$  be a four-vector (in above Arnowitt-Deser-Misner practice $\mu,\nu = 0,1,2,3$ used).

However this is not true, because the counterexample 
\begin{equation}
T^{\mu\nu}=K^{\mu}K^{\nu}
\label{eqd4}
\end{equation}
in Appendix \ref{appc} itself satisfies the dominant energy condition 
\begin{equation}
T^{44}\ge |T^{\mu\nu}|, 
\label{eqd5}
\end{equation}
namely 
\begin{align}
K^4 K^4 \ge |K^{\mu}K^{\nu}| 
\label{eqd6}
\end{align}
or
\begin{align}
(\omega/c)^2 \ge |K^{\mu}K^{\nu}|,
\label{eqd7}
\end{align} \\
where $K^{\mu}=(\mathbf{k}_{\mathrm{w}},\omega/c)$  with $|\mathbf{k}_{\mathrm{w}}|=\omega/c\ge |K^{\mu}|$ (here Rindler's practice $\mu,\nu = 1,2,3,4$ \cite[p.\,138]{r22}  used).  

That is to say, although the conservation law $T^{\mu\nu}_{~~~,\nu}=0$ or $\partial_{\nu}T^{\mu\nu}=\partial_{\nu}(K^{\mu}K^{\nu})=0$  and the dominant energy condition $T^{44}\ge |T^{\mu\nu}|$ or $K^4 K^4 \ge |K^{\mu}K^{\nu}|$  are both satisfied, 
\begin{align}
P^{\mu}:&=\int T^{\mu 4}\mathrm{d}^3x
\notag \\
&=\int K^{\mu}K^4\mathrm{d}^3x
\notag \\
&=\left(\int K^4\mathrm{d}^3x\right)K^{\mu}
\label{eqd8}
\end{align} 
is not a four-vector, because $\big(\int K^4\mathrm{d}^3x\big)$ is not a scalar, as shown in Eq.\,(\ref{eqc5}) of Appendix \ref{appc}.\\
\indent From above we can see that all the proofs of the positive mass theorem \cite{r32,r33,r34,r35,r36,r39} are based on a flawed theoretical framework, and thus the validity of the theorem itself could be called into question. 

\section{\label{appe} Is Gordon optical metric covariant?}
How to define the covariance of physical quantities/equations is one of the most important issues in optics of moving media. Actually, it is a long-lasting unsolved fundamental problem in the theory of relativity. In this appendix, we take Maxwell equations and the four-momentum of a massive particle as examples to give the definition of covariance, and indicate why the covariance of Gordon metric tensor \cite{r42} is indeterminate.    

In principle, the definitions of physical quantities used to construct a covariant theory of light propagation in moving media are supposed to be given in a general inertial frame, otherwise the covariance of the constructed theory is ambiguous. At least, one cannot identify  the covariance of the quantity without a definition given in a general frame.

As a first principle, Maxwell equations are the physical basis for descriptions of macro electromagnetic phenomena, and all other alternative approaches, including Lagrange formalism, must be confirmed by the Maxwell equations \cite[p.\,598]{r4}.  Thus any justified results that cannot be derived from Maxwell equations could constitute a challenge of the completeness of classical electromagnetic theory. 

It is well-established that Maxwell equations are covariant under Lorentz transformation of two EM field-strength tensors \cite{r24}.  This covariance has three properties:
\begin{enumerate}
\item [(a)] Maxwell equations keep invariant in mathematical form in all Lorentz inertial frames.
\item[(b)] All physical quantities appearing in the equations have the same physical definitions in all frames.
\item[(c)] Maxwell equations in one frame do not include any physical quantities defined in other inertial frames.
\end{enumerate}
The above three properties do not contain each other. To put it simply, Maxwell equations are \emph{frame-independent} under Lorentz transformation of two EM field-strength tensors.

Another typical example for the covariance is the four-momentum of a massive particle, given by \\
\begin{align}
&~~~~m_0U^{\mu}~~~~~~~~~~~\,\textrm{(Non-covariant form)}
\notag \\
\notag \\
&=\frac{m}{\gamma}U^{\mu}~~~~~~~~~~~~\textrm{(Covariant form in relativity)}
\notag \\ 
&= \left(m\mathbf{v},\frac{mc^2}{c}\right)~~\,\textrm{(Covariant~form~in~3D space)}
\label{eqe1} 
\end{align} \\
in a general $XYZ$ frame, where $m_0$  is the particle's mass in the \emph{particle-rest frame}\,---\,rest (proper) mass, $U^{\mu}=\gamma(\mathbf{v},c)$  is the four-velocity,  $m\,(=\gamma m_0)$ is the mass, $\gamma=(1-\mathbf{v}^2/c^2)^{-1/2}$, and $\mathbf{v}$  is the particle's velocity.  

The definition of $m_0U^{\mu}$ is in a non-covariant form. $m_0U^{\mu}$ has the same mathematical form and physical definition in all inertial frames, and complies with Properties (a) and (b), but it is not consistent with Property (c), because  $m_0U^{\mu}$  includes $m_0$ defined in the particle-rest frame, and thus  $m_0U^{\mu}$  cannot provide the frame-independent definitions of momentum and mass in a general frame. For example, $m_0U^{\mu}$  defines $m_0$ as the particle's mass in the particle-rest frame, but  $m_0U^{\mu}$  does not define the particles's mass $m$ in a general frame.

Properties (b) and (c) require that the definitions of particle's momentum and mass must be the same in all inertial frames and they do not include any quantities defined in other frames, namely
\begin{equation}
\textrm{Momentum}=\textrm{Mass}\times \textrm{Velocity}=m\mathbf{v}.
\label{eqe2}
\end{equation}
Thus we have the mass transformation  $m=\gamma m_0$.

Obviously, the definition  of four-momentum $(m/\gamma)U^{\mu}=(m\mathbf{v},mc^2/c)$ is frame-independent under its Lorentz transformation.  Especially, like Maxwell equations, it does not include any quantities of other frames; thus this definition complies with Property\,(c), in addition to Properties (a) and (b).

Note that $(m/\gamma)U^{\mu}$  is a covariant form in relativity, with $(m/\gamma)$ an invariant and   $U^{\mu}$ the four-velocity of the particle in a general $XYZ$ frame, while $(m\mathbf{v},mc^2/c)$  is a covariant form in 3D-space, with $m\mathbf{v}$  the momentum and $mc^2$  the energy of the particle in a general $XYZ$ frame; just like Maxwell equations have the two forms of covariance (confer \cite[Footnote 7 there]{r17}).

Unfortunately, Property\,(c) has been usually neglected in constructing a tensor in the community.  A typical example is Gordon optical metric tensor \cite{r42}, which is generalized to describe the effective gravitational field for a nonuniformly moving medium in the works by Leonhardt and Piwnicki \cite{r43,r44}.  The Gordon metric in a general $XYZ$ frame reads:
\begin{equation}
\mathnormal{\Gamma}^{\mu\nu}=g^{\mu\nu}+(n_0^2-1)u^{\mu}u^{\nu},
\label{eqe3}
\end{equation}
where $g^{\mu\nu}=\mathrm{diag}(-1,-1,-1,+1)$  is the Minkowski metric, $u^{\mu}=U^{\mu}/c=\gamma (\mb{\beta},1)$  with $\mb{\beta}=\mathbf{v}/c$  is the normalized four-velocity of the moving medium (namely the medium moves at $\mathbf{v}=\mb{\beta}c$  with respect to $XYZ$), and $n_0$  is the refractive index in the \emph{medium-rest frame}.

From Eq.\,(\ref{eqe3}), we find that $\mathnormal{\Gamma}^{\mu\nu}=g^{\mu\nu}+(n_0^2-1)u^{\mu}u^{\nu}$  is given in a general frame $XYZ$, but it includes $n_0$  which is a quantity defined in another frame (medium-rest frame).  Obviously, this is not consistent with Property\,(c).  In their works \cite{r43,r44}, Leonhardt and Piwnicki are supposed to provide the definition of the refractive index in a general frame in order to remove this inconsistency and make it expressed in a covariant form, just like Eq.\,(\ref{eqe2}) leading to the \emph{frame-independent} four-momentum  $(m/\gamma)U^{\mu}=(m\mathbf{v},mc^2/c)$  shown above; unfortunately, the authors failed to do so.  Thus the covariance of  $\mathnormal{\Gamma}^{\mu\nu}$  cannot be identified, namely the covariance of Gordon optical metric $\mathnormal{\Gamma}^{\mu\nu}$ is ambiguous, depending on how to define the refractive index in a general inertial frame.\\
\indent It should be indicated that Gordon defined the (proper) refractive index $n_0$  by assuming that the medium is isotropic, observed in the medium-rest frame \cite{r42}.  However when a medium is moving, it becomes anisotropic in general \cite{r45}, and Gordon's way to define $n_0$  is not valid to define the refractive index in a general inertial frame. \\
\indent It also should be indicated that the way to define the refractive index in a general frame could essentially revise the descriptions of light propagation.  For example, in Wang's theory where the refractive index in a uniform medium is defined in a general inertial frame, propagation of light complies with Fermat's principle in all inertial frames \cite{r7}, while in Leonhardt's theory, it does only ``in the special case of a medium at rest'' \cite{r46}.


\newpage

\begin{figure} 
\includegraphics[trim=1.0in 1.0in 1.0in 1.0in, clip=true,scale=1.0]{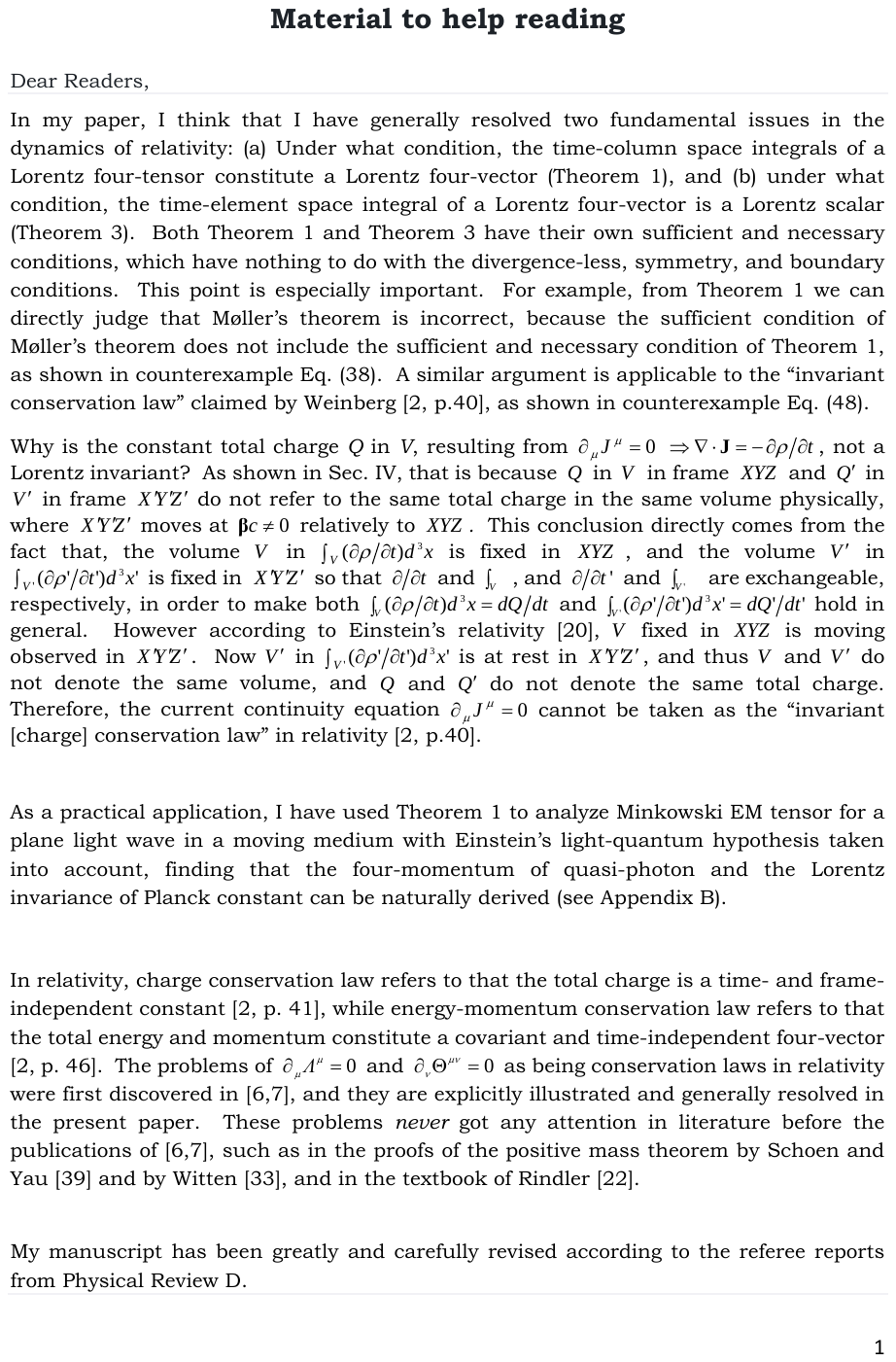}
\label{figM1}
\end{figure} 

\newpage

\begin{figure} 
\includegraphics[trim=1.0in 1.0in 1.0in 1.0in, clip=true,scale=1.0]{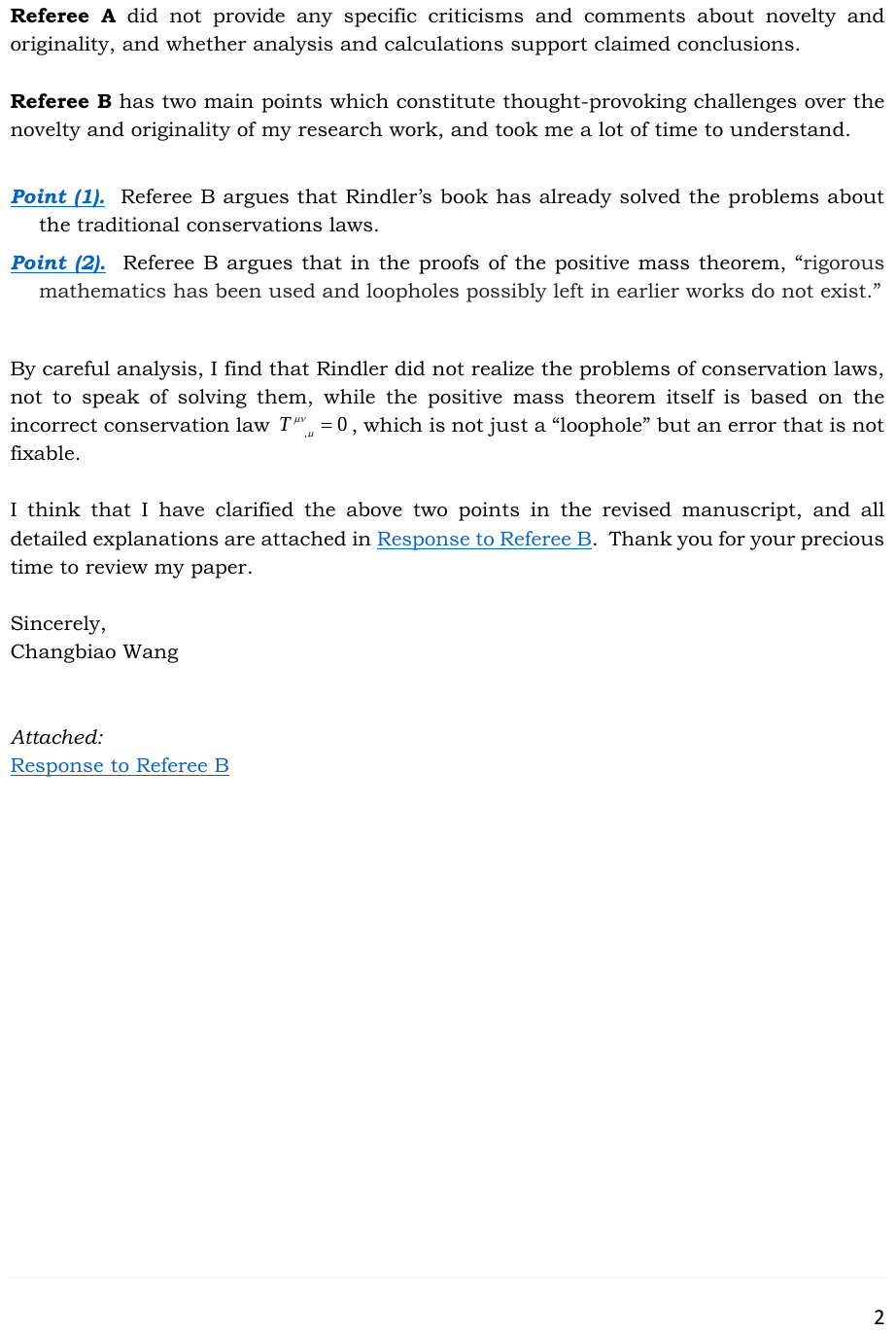}
\label{figM2}
\end{figure} 

\newpage

\begin{figure} 
\includegraphics[trim=1.0in 1.0in 1.0in 1.0in, clip=true,scale=1.0]{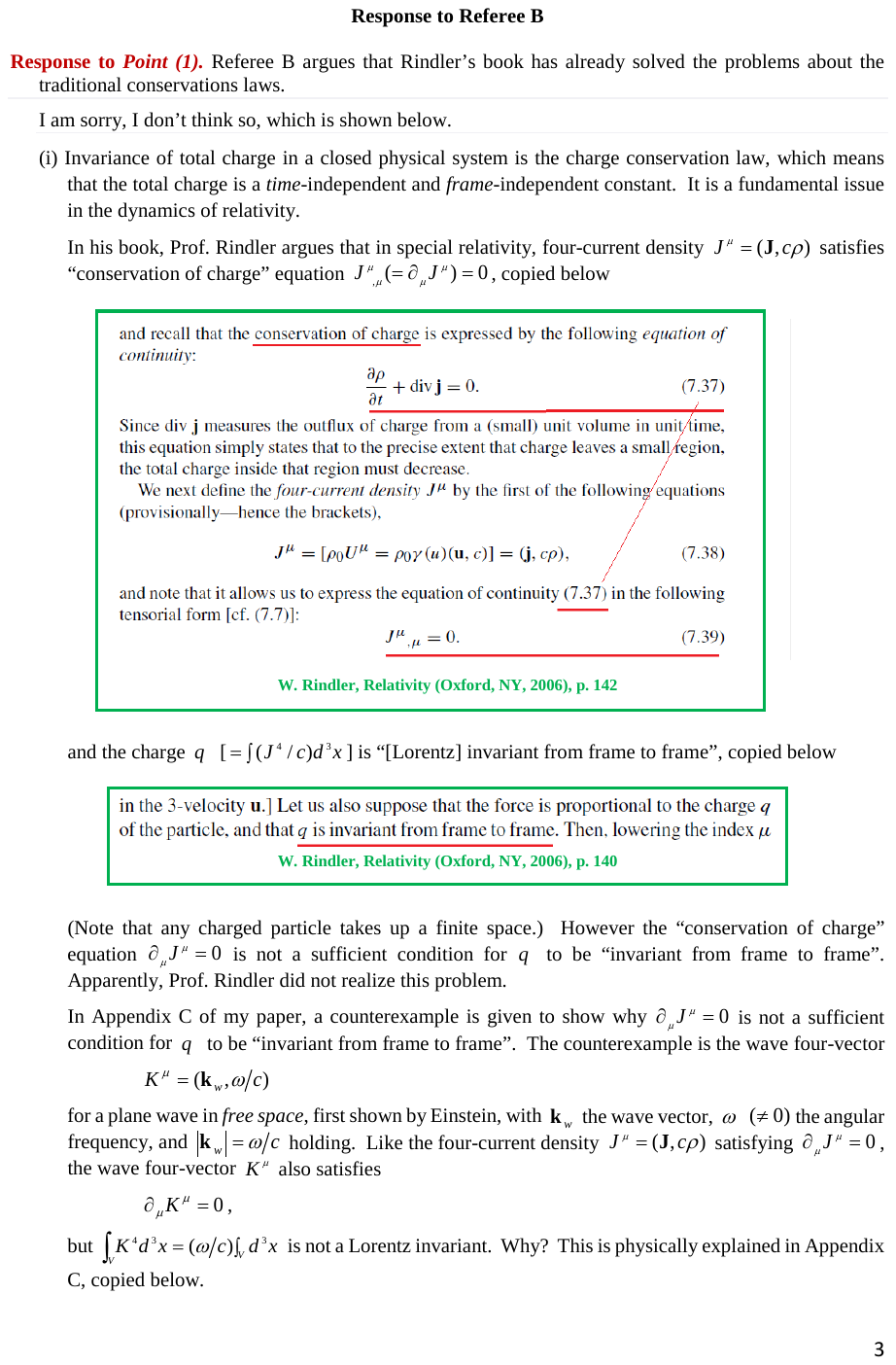}
\label{figM3}
\end{figure} 

\newpage

\begin{figure} 
\includegraphics[trim=1.0in 1.0in 1.0in 1.0in, clip=true,scale=1.0]{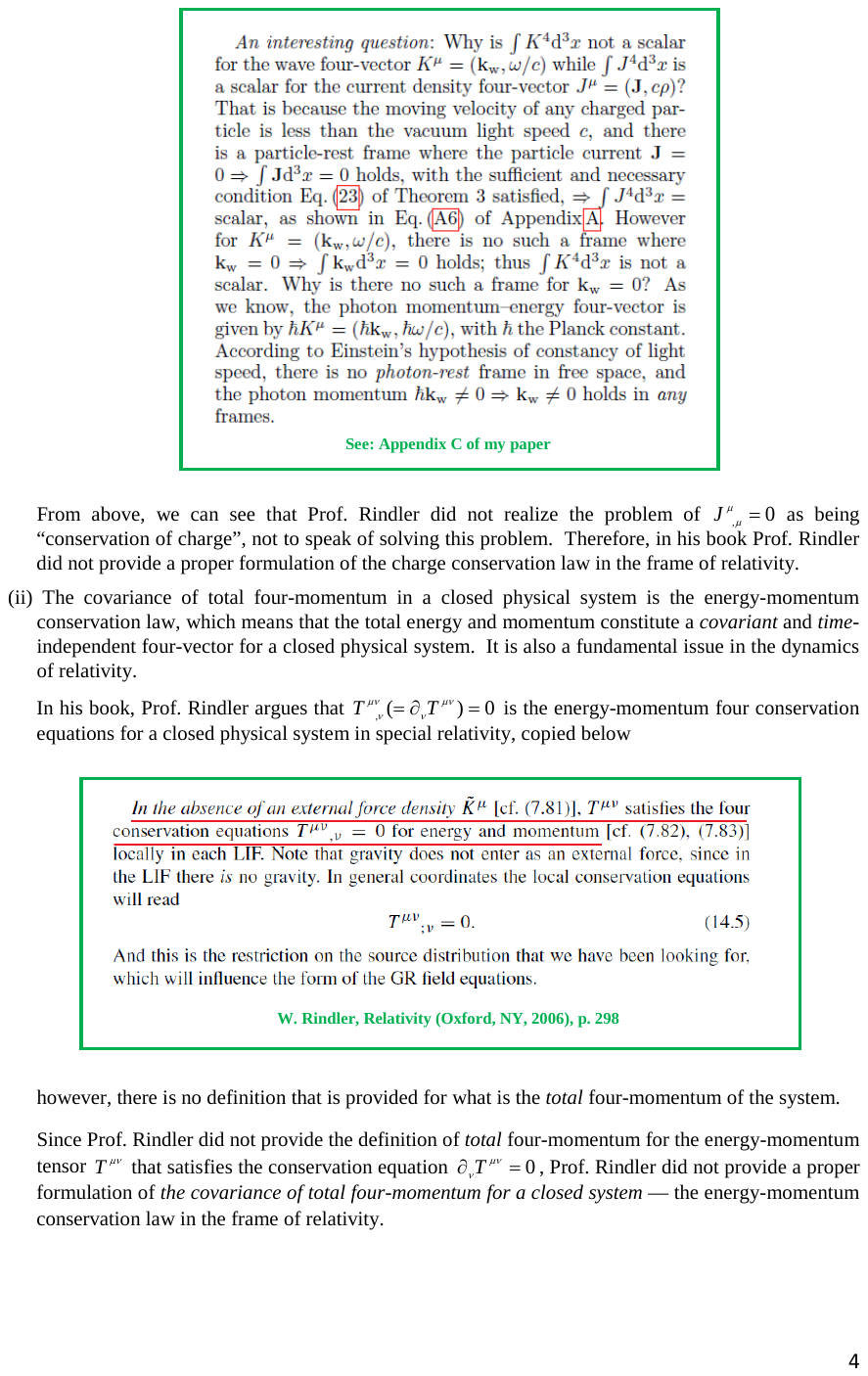}
\label{figM4}
\end{figure} 

\newpage

\begin{figure} 
\includegraphics[trim=1.0in 1.0in 1.0in 1.0in, clip=true,scale=1.0]{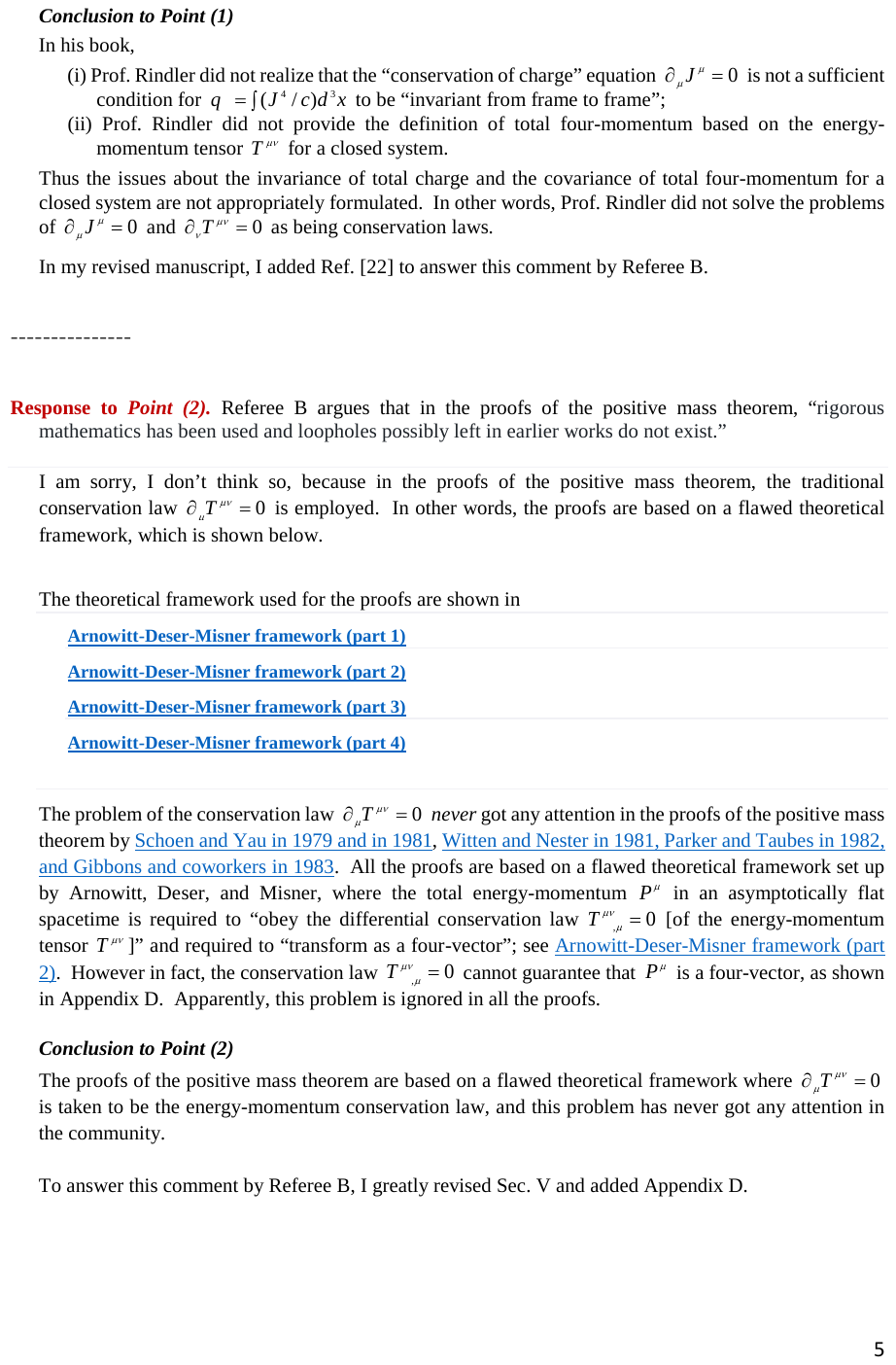}
\label{figM5}
\end{figure} 

\newpage

\begin{figure} 
\includegraphics[trim=1.0in 1.0in 1.0in 1.0in, clip=true,scale=1.0]{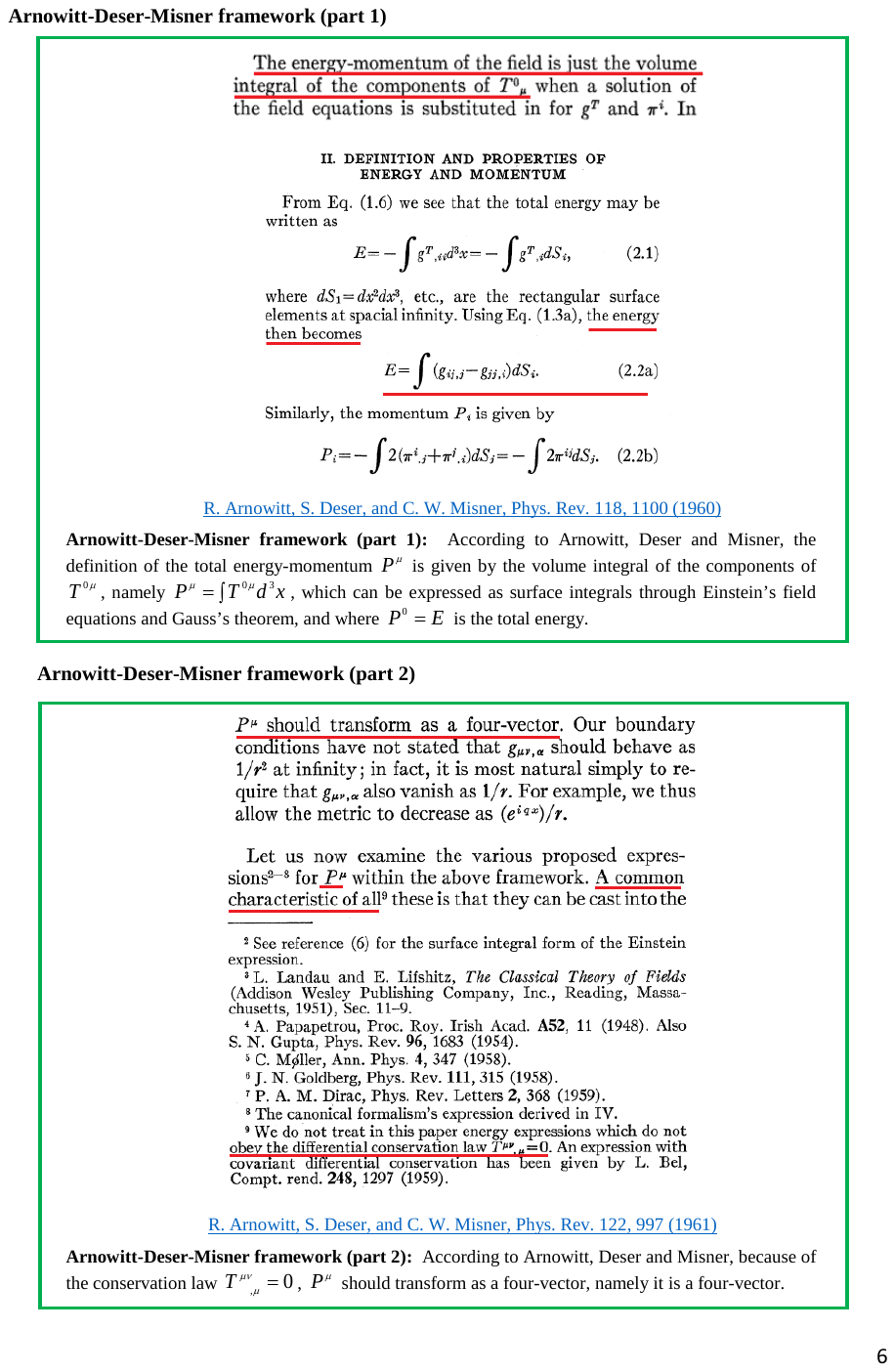}
\label{figM6}
\end{figure} 

\newpage

\begin{figure} 
\includegraphics[trim=1.0in 1.0in 1.0in 1.0in, clip=true,scale=1.0]{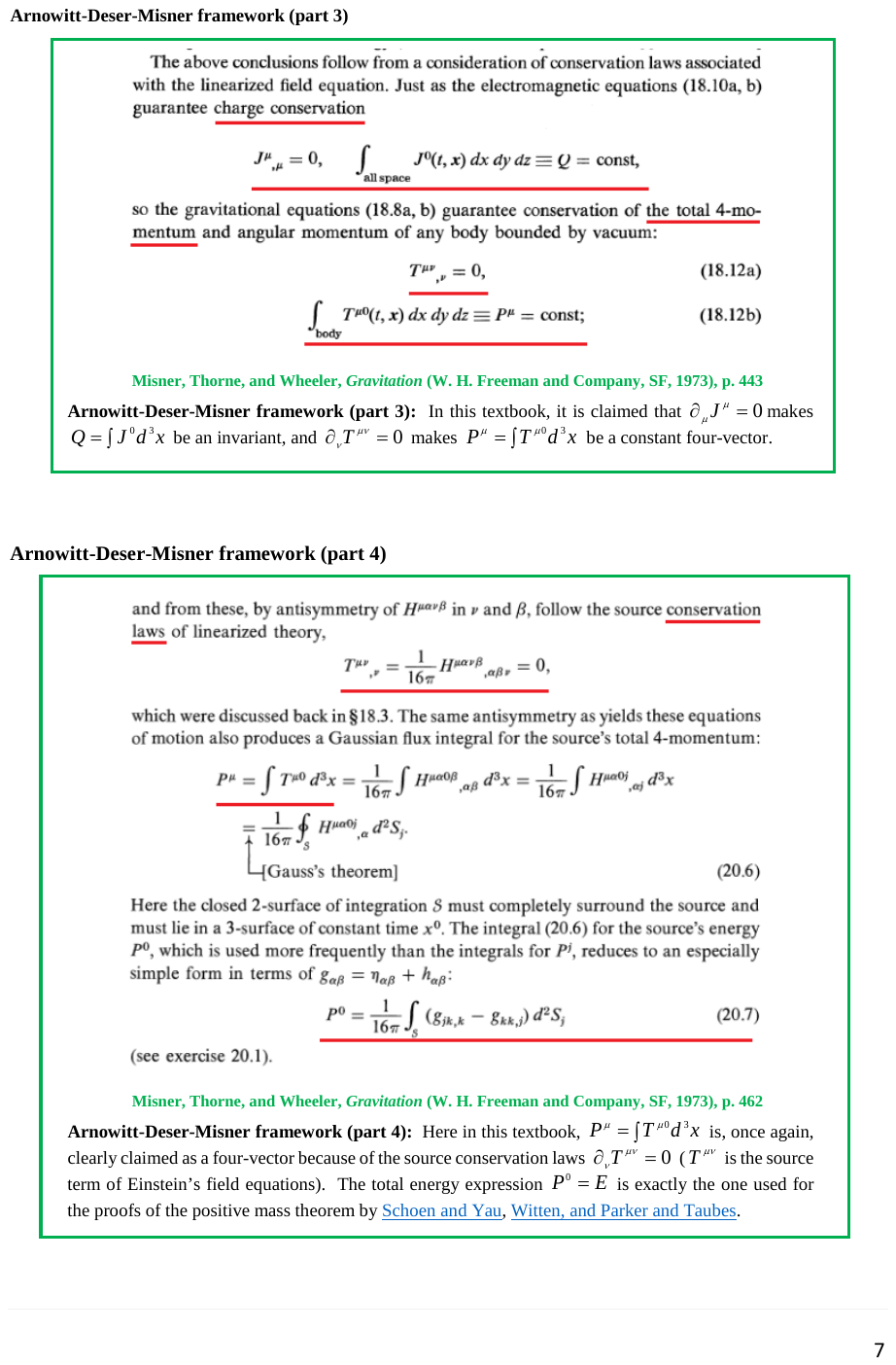}
\label{figM7}
\end{figure} 

\newpage

\begin{figure} 
\includegraphics[trim=1.0in 1.0in 1.0in 1.0in, clip=true,scale=1.0]{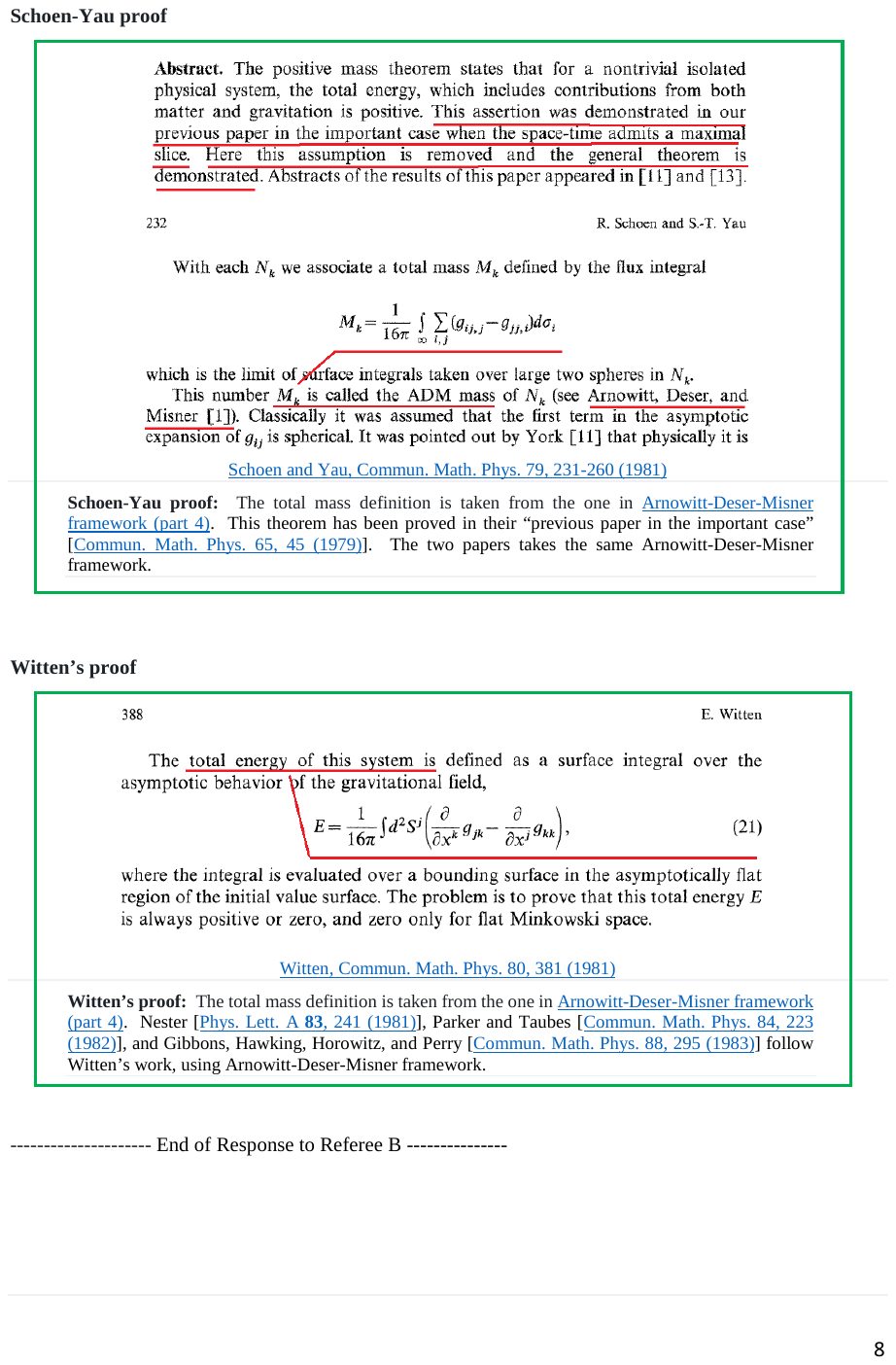}
\label{figM8}
\end{figure} 

\end{document}